\documentclass[review,fleqn,sort&compress]{elsarticle}
\usepackage{amssymb,amsbsy,amsthm,amsmath,amsfonts,amssymb,amscd}
\usepackage{geometry}
\usepackage{bm} 
\usepackage{graphicx}
\usepackage{float}
\usepackage{multirow}
\usepackage{subfigure}
\usepackage{lineno}
\usepackage{blindtext}
\usepackage{hyperref}
\hypersetup{colorlinks, citecolor=black, filecolor=black, linkcolor=black, urlcolor=black}
\usepackage{booktabs}
\usepackage[ruled,linesnumbered]{algorithm2e}
\usepackage{appendix}
\usepackage{algpseudocode}
\usepackage{mathtools,nccmath}
\usepackage{adjustbox}
\usepackage{cleveref}
\usepackage{arydshln}
\usepackage{tikz}
\usepackage{hhline}
\usetikzlibrary{shapes, arrows, positioning}
\usepackage{dblfloatfix}
\usepackage{url}
\usepackage{pythonhighlight}

\usepackage{listings}
\definecolor{dkgreen}{rgb}{0,0.6,0}
\definecolor{gray}{rgb}{0.5,0.5,0.5}
\definecolor{mauve}{rgb}{0.58,0,0.82}

\graphicspath{{fig/}}
\theoremstyle{definition}

\usepackage{color}
\usepackage{soul}

\begin{document}

\begin{frontmatter}

\title{Image-based adaptive domain decomposition for continuum damage models}

\author[1]{Panos Pantidis\corref{cor}}
\author[2]{Cornelius Otchere}
\author[1]{Mostafa E. Mobasher}

\address[1]{Civil and Urban Engineering Department, New York University Abu Dhabi, Abu Dhabi, P.O. Box 129188, UAE}
\address[2]{Department of Civil and Systems Engineering, Johns Hopkins University, Baltimore, MD}



\cortext[cor]{Corresponding author. \emph{E-mail address:} \texttt{pp2624@nyu.edu} (Panos Pantidis)}

\begin{highlights}

\item We propose an innovative approach to speed up nonlinear FEM computations using images. 

\item A low-cost and colormap-independent scheme is developed to reliably detect and track damage features.

\item The detected damage contours inform a domain decomposition scheme, and time savings stem from the Schur complement.

\item Computational savings scale with model size, and here we report a speedup up to 4 times than single-domain analysis.

\item Since the images are problem-agnostic, this framework can be easily extended to other computational models.

\end{highlights}

\begin{abstract}

We present a novel image-based adaptive domain decomposition FEM framework to accelerate the solution of continuum damage mechanics problems. The key idea is to use image-processing techniques in order to identify the moving interface between the \textit{healthy subdomain} and \textit{unhealthy subdomain} as damage propagates, and then use an iterative Schur complement approach to efficiently solve the problem. The implementation of the algorithm consists of several modular components. Following the \textit{FEM solution} of a load increment, the \textit{damage detection} module is activated, a step that is based on several image-processing operations including colormap manipulation and morphological convolution-based operations. Then, the \textit{damage tracking} module is invoked, to identify the crack growth direction using geometrical operations and ray casting algorithm. This information is then passed into the \textit{domain decomposition} module, where the domain is divided into the healthy subdomain which contains only undamaged elements, and the unhealthy subdomain which comprises both damaged and undamaged elements. Continuity between the two regions is restored using penalty constraints. The computational savings of our method stem from the Schur complement, which allows for the iterative solution of the system of equations appertaining only to the unhealthy subdomain. We perform an exhaustive comparison between our approach and single domain computations through a series of benchmark examples, and we demonstrate the accuracy, efficiency, and numerical robustness of the proposed framework. We ensure the flexibility of our method by testing it on both local and non-local integral damage laws and structured and unstructured mesh idealizations, and we extend it to more challenging problems where different damage paths eventually merge. Since the key novelty of the method lies in using image processing tools to inform the domain decomposition module, our framework can be readily extended beyond damage mechanics and model several classes of non-linear problems such as plasticity, phase-field, and more. The code and data used in this work will be made publicly available upon publication of the article.

\end{abstract}

\begin{keyword}
\texttt image-processing \sep damage mechanics \sep domain decomposition \sep non-linear problems \sep finite element method
\end{keyword}

\end{frontmatter}

\newpage


\section{Introduction}
\label{Sec:Introduction}

\subsection{Thesis statement}

Many computational models, including damage and fracture models, are notorious for exhibiting elevated computational costs. In both literature and practice, several methods have been developed to counteract this amplified computational expense and promote the solution efficiency. However, many of these methods are either difficult to implement, or specific to one class of models. In this work, we propose an enhanced solution approach that relies on contour images, which are practically produced by almost every modern solver for visualization and post-processing purposes. In this study we focus on continuum damage modeling. The idea is to employ adaptive domain-decomposition and a Schur complement based solver in order to speed up the solution by applying the iterative non-linear solution only on the \textit{unhealthy subdomain}, while the stiffness of the \textit{healthy subdomain} remains constant. The novelty of this study relies on implementing a suite of image processing techniques to automatically and adaptively detect and track damage, and then advise the geometries of the subdomains.

\subsection{Literature review}
The propagation of cracks and voids inside the microstructure of a material is highly detrimental for its macroscopic mechanical properties, and consequently it can pose a serious threat to the overall structural integrity. Understanding and mitigating fracture propagation both at the material and at the structural level is a critical prerequisite for establishing safe design protocols across many engineering fields \cite{li2011ductile, kanvinde2017predicting, zimmermann2020review, olson2008multi}, and a plethora of theoretical approaches have been proposed over the last decades to describe fracture-associated phenomena \cite{griffith1921vi, perez2004linear, kachanov1986introduction, francfort1998revisiting, lemaitre2006engineering, wu2020phase, ha2011characteristics, moes2021lipschitz, jirasek1998nonlocal, volegov2016damage, de2016gradient, diehl2022comparative}. In light of the elevated expense entailed by experimentally assessing fracture-induced failures, the main pathway by which these frameworks are evaluated is computational fracture modeling. Several mature frameworks exist on this front, such as the Finite Element Method (FEM) \cite{kuna2013finite, song2008comparative}, the eXtended Finite Element Method (XFEM) \cite{sukumar2000extended, jiang2021xfem}, the Virtual Element Method (VEM) \cite{benedetto2014virtual, artioli2020vem}, the Boundary Element Method (BEM) \cite{cruse2012boundary, chen2021modeling}, meshless approaches \cite{rao2000efficient}, and more \cite{de2011x, borden2014higher}. Undoubtedly, the robust development of these frameworks has significantly advanced the research frontiers and has enabled the understanding and exploration of complex fracture mechanisms. 

Despite the significant progress, a commonly shared bottleneck of most numerical frameworks is their simulation expense. Particularly for FEM, this is a byproduct of the need for a fine mesh discretization with finite elements, especially around the crack tip and along the fracture path. The fine mesh is required to capture accurately the crack initiation and the correct crack propagation trajectory. The computational cost is significantly amplified if one includes additional layers of complexity, such as the presence and interaction of multiple cracks, material anisotropy, multi-phase materials, large-scale domains spanning multiple length scales, or 3D considerations \cite{nguyen2020adaptive, ferte20163d, nguyen2017multi, zulian2021large, giovanardi2020fully}. In particular, several studies report simulation times that can take several hours, days or even weeks to complete \cite{pantidis2024fenn, nguyen2020adaptive, nguyen2017multi, feng2023past}. This evidently underlines the importance of developing robust methods to accelerate the numerical solution of such problems, in order to enable their seamless utilization for industrial-level purposes. In order to counteract the notorious expense of fracture computational models researchers have explored a variety of advanced modeling approaches, and these are briefly discussed below.

One major pathway to reduce this computational cost includes domain decomposition (DD) frameworks. The core idea of DD approaches is to split a large and complex domain into several smaller subdomains, solve each one separately, and then combine the individual solutions into a global one while satisfying necessary continuity constraints at the interfaces of the subdomains. Several notable DD frameworks for fracture modeling exist, such as the Finite Element Tearing and Interconnecting (FETI) approach \cite{farhat1991method} and its successors \cite{kozubek2013total, gosselet2015simultaneous, xu2021feti}, space-time domain decomposition with mixed formulations \cite{hoang2016space}, parallel computing using DD pre-conditioners, \cite{svolos2020updating}, and more \cite{bakalakos2022domain}. Moving beyond domain decomposition, major efforts have also been made to enhance currently available numerical techniques by implementing advanced algorithmic variations in the conventional approaches. Without being nearly exhaustive, we mention modifications to the Newton-Raphson \cite{sepasdar2020overcoming} and the arc-length approach \cite{saji2024ual}, GPU parallelization and multi-thread optimization \cite{zulian2021large, zhang2024improved}, the development of staggered schemes \cite{liu2016abaqus, brun2020iterative} and coupled methods \cite{mobasher2016adaptive, tao2011fully}, as well as newly-emerged frameworks rooted on neural networks \cite{pantidis2024fenn, goswami2022physics}. Overall, as the size, scale and complexity of the fracture problems that are sought to be investigated increase, there is a growing need for developing new approaches in this already active field of research.

More recently, researchers started exploring the potential of using image-processing techniques to aid, complement or even bypass the computational modeling of fracture-related problems. Arguably, a research path where image-processing tools are particularly efficient, is the detection of cracks and other discontinuities inside the domain. Several methodologies with varying levels of sophistication have been proposed on this front, and the interested reader is referred to \cite{munawar2021image, gupta2022image} for comprehensive reviews of image-based techniques focusing on crack detection. A closely related and currently growing field involves studies on image-based prediction of crack propagation, and we mention, for example \cite{pantoja2022determining}, where the authors proposed a methodology to determine the crack kinematics for Mode I and Mode II problems based on image-crack patterns. Nevertheless, all the aforementioned approaches share a common conceptual basis: images are utilized to characterize the topological arrangement (static or evolving) of the crack network. A conceptually different approach of utilizing image-based techniques in fracture modeling regards the identification of constitutive laws. For example, \cite{mathieu2013image} proposed an image-based process to identify a crack propagation law, and \cite{broggiato2007identification} developed an inverse approach using Digital Image Processing to identify the parameters of a material damage model. Images have found several other uses, see for example, \cite{wang2024image}, where the authors developed a convolution-based method to automate the irregular stone packing in the construction of masonry walls, and \cite{trent2023using}, where the authors trained a Generative Adversarial Network on image-pairs containing state variables (stresses, displacements, etc.) to emulate the entire FEM simulation. Altogether, it is evident that image-based algorithms are currently utilized in several ways as complementary methods to conventional computational modeling, and the development of more sophisticated machine learning algorithms and image-processing tools is expected to further accelerate the growth of this field. 

\subsection{Scope and Outline}

The contribution of this article is the development of an image-based adaptive domain decomposition framework that accelerates the numerical solution of damage mechanics problems. The major novelty of our work lies in using image-analysis techniques to monitor the propagation of damage and adaptively update the domain decomposition component of our framework. Conceptually this approach is clearly distinct from the relevant literature, and to the best of our knowledge this is the first time that image-processing methods are utilized in this fashion. To achieve our goal, we focus on continuum damage mechanics and we devise a framework which comprises the following four modules: a) FEM solver, b) damage detection, c) damage tracking and d) domain decomposition. The first module is invoked to perform the non-linear analysis of a load increment, at the end of which we produce an image of the damage contours. Then, we deploy a series of image-recognition algorithms whose purpose is to detect the presence of damage inside the domain. If damage is not detected we continue with the analysis of the next load increment. Otherwise, we proceed to the third module, and we perform an appropriate offset-scaling process to identify the FEM nodes that are enclosed by the area with the detected damage. Finally, we pass this information to the penalty-based domain decomposition module, and utilize it to separate the domain into a healthy subdomain (no damage) and an unhealthy subdomain (with damage). This allows us to arrive at the final partitioned system of equations, which we solve efficiently using the Schur complement method. In this study we focus on the application of the framework to damage mechanics problems, and we examine several benchmark numerical examples using both the local and the non-local integral damage models. However, we note that our method can be readily extended to other non-linear models such as plasticity or phase-field, or even be extended to process other types of images such as ultrasound, X-ray, and micro-CT. Compared to the conventional single domain analysis we record computational savings that scale with the model size, and we find that our method can overcome severe numerical instabilities that hinder the convergence of the conventional solver. 

The paper is organized as follows. Section \ref{Sec:CDM} contains a brief presentation of the continuum damage mechanics models adopted in this study. Section \ref{Sec:Methodology} illustrates the overarching structure of the proposed framework, and Sections \ref{Sec:FEM_Implementation} - \ref{Sec:Domain_Decomposition} present a detailed description of the utilized sub-routines. The results of the framework's numerical implementation are presented in Section \ref{Sec:Numerical_Results}, followed by the summary and conclusions in Section \ref{Sec:Summary_Conclusions}.

\section{Continuum Damage Mechanics (CDM) models}
\label{Sec:CDM}

In this section, we lay out the mathematical background of the two CDM theories that we apply in our numerical examples: the \textit{local damage} method and the \textit{non-local integral damage} approach. Scalar values are denoted with regular-font letters (e.g., $d$), first- and second-order tensors are denoted with bold letters (e.g., $\boldsymbol{\sigma}$), and fourth-order tensors with black-board bold capital letters (e.g. ${\mathbb{C}}$).

\subsection{Local damage model}
\label{Sec:Local_CDM}

\begin{figure}
    \centering
    \includegraphics[width=0.5\textwidth]{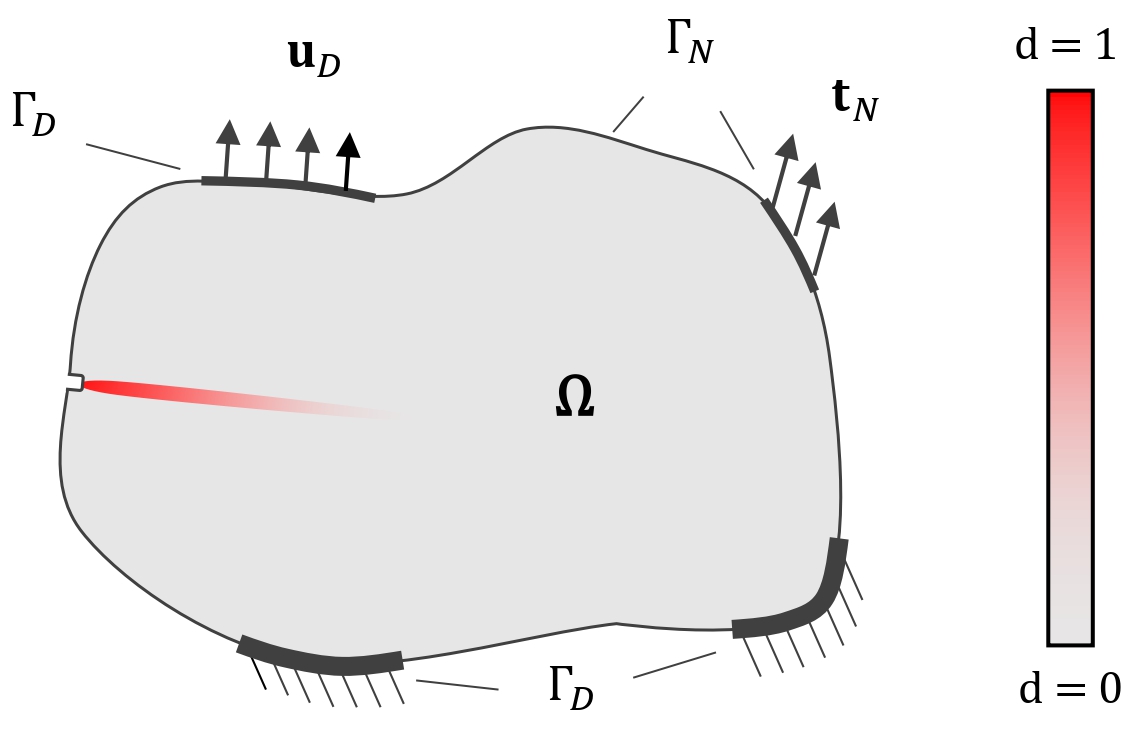}
    \caption{Generic schematic view of a domain with a sample damage contour.}
    \label{Figure_elastic_domain}
\end{figure} 

Consider an isotropic domain $\Omega$ with Dirichlet boundary $\Gamma_{D}$ and Neumann boundary $\Gamma_{N}$, as shown in Fig. \ref{Figure_elastic_domain}. In the absence of body forces, the equilibrium equation reads:

\begin{equation}
    \boldsymbol{\nabla} \cdot \boldsymbol{\sigma} = 0 \quad in \quad \Omega 
\label{Eqn_equilibrium_1}
\end{equation}

accompanied by the following boundary conditions:

\begin{equation}
	{\boldsymbol{\sigma}} \cdot {\boldsymbol{n}} = {\boldsymbol{t}}_{N} \quad on \quad \Gamma_{N} \; ; \; \; \;  
    {\boldsymbol{u}} = {\boldsymbol{u}}_{D} \quad on \quad \Gamma_{D}
\label{Eqn_equilibrium_2}
\end{equation}

In the expressions above, ${\boldsymbol{\sigma}}$ is the Cauchy stress, ${\mathbb{C}}$ is the elasticity tensor, ${\boldsymbol{\varepsilon}}$ is the strain tensor, ${\boldsymbol{n}}$ is the unit outward vector, ${\boldsymbol{t}}_{N}$ is the imposed traction and ${\boldsymbol{u}}_{D}$ is the prescribed displacement loading. Following the CDM paradigm and the strain equivalence assumption \cite{kachanovbook}, a damage variable $d$ that represents the stiffness deterioration is introduced into the material constitutive law. This is a scalar variable which ranges as $d: 0\leq d\leq 1$, with $d = 0$ representing the undamaged state and $d = 1$ the complete material failure. The expression of the effective stress is given as:

\begin{equation}
    {\boldsymbol{\sigma}}_{d} = (1 - d) \; {\mathbb{C}} \; : {\boldsymbol{\varepsilon}}
\label{Eqn_equilibrium_3}
\end{equation}

The damage variable $d$ is typically computed as a function of the equivalent strain $\varepsilon_{eq}$ ($d = d(\varepsilon_{eq})$), which is a macroscopic measure of the local material point deformation \cite{lemaitre2006engineering}. The local damage framework is very efficient from a computational standpoint because $d$ is computed independently at any given material point, without the need of additional degrees of freedom \cite{peerlings1996gradient} or spatial averaging \cite{bazant2002nonlocal}. However, its numerical implementation leads to well-documented spurious results, due to the loss of ellipticity in the underlying PDE \cite{peerlings1996gradient,pijaudier1987nonlocal}. Essentially damage is pathologically localized at a single element, and fracture propagation becomes dependent on the mesh discretization. This drawback is overcome if one adopts a non-local method. Below we discuss one such approach, the non-local integral model \cite{pijaudier1987nonlocal}.

\subsection{Non-Local integral damage model}
\label{Sec:Non_Local_Integral_CDM}

There is a plethora of non-local models in the literature, such as the non-local gradient \cite{peerlings1996gradient}, lip-field \cite{moes2021lipschitz}, etc. \cite{wu2020phase}, and the interested reader is referred to \cite{jirasek1998nonlocal} for a detailed review. The primary objective of all these methods is the regularization of the local damage field, in order to alleviate the mesh sensitivity of the numerical solution. In this study we implement the non-local integral damage formulation \cite{pijaudier1987nonlocal}. According to this framework, a non-local damage variable $\bar{d}$ is calculated at any material point ${\bf{X}}^{i}_{g}$ as the weighted average of the local damage of all points ${\bf{X}}_{g}$ that lie within a specified radial distance from ${\bf{X}}^{i}_{g}$. The non-local damage variable $\bar{d}({\bf{X}}^{i}_{g})$ is computed as:

\begin{equation}
    \bar{d}({\bf{X}}^{i}_{g}) = \frac{\int_\Omega \Phi({\bf{X}}^{i}_{g}, {\bf{X}}_{g}) d({\bf{X}}_{g})\, d\Omega}{\int_\Omega \Phi({\bf{X}}^{i}_{g}, {\bf{X}}_{g}) \, d\Omega}
\label{Eqn_nonlocal_integral_damage}
\end{equation}

\noindent where $\Phi$ is a weighting function expressed as \cite{pijaudier1987nonlocal, mobasher2016adaptive}:

\begin{equation}
    \Phi({\bf{X}}^{i}_{g}, {\bf{X}}_{g}) = exp \left( -\frac{\|{\bf{X}}^{i}_{g} - {\bf{X}}_{g}\|^2}{l_{c}^2} \right)
\label{Eqn_nonlocal_integral_bell_curve}
\end{equation}

\begin{figure}[H]
    \centering
    \includegraphics[width=0.85\linewidth]{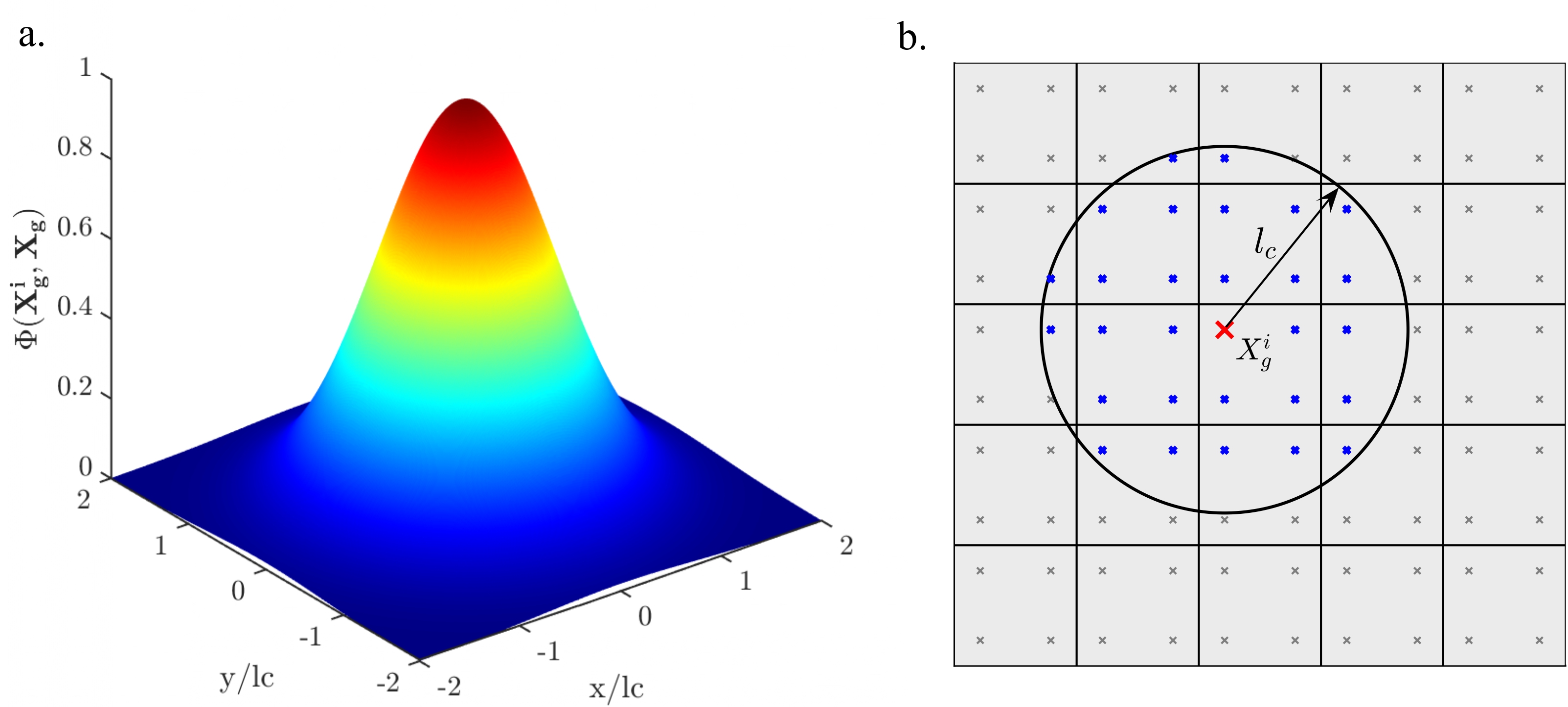}
    \caption{Schematic details of the non-local integral damage method. {\bf{a.}} Bell-shaped surface denoting the weight function $ \Phi({\bf{X}}^{i}_{g}, {\bf{X}}_{g})$ {\bf{b.}} Schematic diagram of Gauss Points ${\bf{X}}_{g}$ within the characteristic length distance $l_{c}$ from a given material point ${\bf{X}}_{g}^{i}$.}
    \label{Figure_Integral_law}
\end{figure}

The $\Phi$ function is a bell-shaped kernel, as shown in Fig. \ref{Figure_Integral_law}a, and attributes more significance to the Gauss points that are closer to ${\bf{X}}^{i}_{g}$. The term $l_{c}$ in Eqn. \ref{Eqn_nonlocal_integral_bell_curve} is the \textit{characteristic length} of the material, and it denotes the length-scale over which damage is regularized. A schematic view of the latter is shown in Fig. \ref{Figure_Integral_law}b. We also remark that in the limiting case of $l_{c} \rightarrow 0$, the non-local damage $\bar{d}$ reduces to its local counterpart at that location. 

\section{Methodology}
\label{Sec:Methodology}

\begin{figure}
    \centering
    \includegraphics[width=1\linewidth]{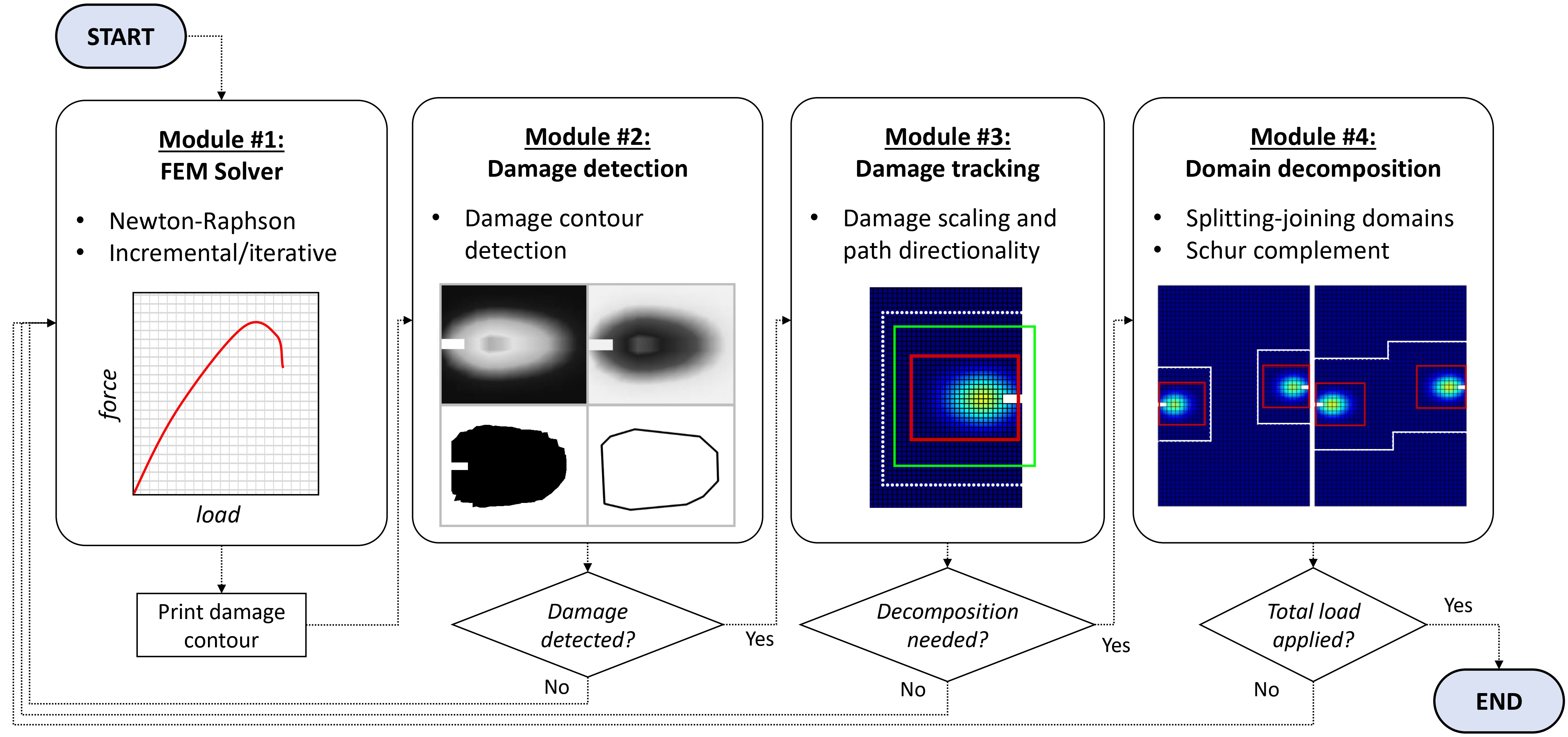}
    \caption{Schematic workflow of the proposed four-module framework.}
    \label{Figure_framework_workflow}
\end{figure}

In this work, we develop the image-based adaptive domain decomposition scheme to solve the boundary value problem formulated in the previous section. For this purpose, we devise a highly modular framework that addresses all the underlying processes. The latter can be classified into four main modules: a) FEM solver scheme, b) damage detection sub-routine, c) damage tracking sub-routine, and d) domain decomposition algorithm. A general flowchart of the proposed methodology is shown in Figure \ref{Figure_framework_workflow}. Below, we present a general overview of the different modules, elaborating on how they are interconnected and explaining their functionality and input-output variables. The specific conceptual and implementation details of each module will be presented in the following sections.

\begin{enumerate}

    \item \textbf{Finite Element implementation}: In this module, we perform the numerical solution using the finite element method and a typical non-linear solver, such as the Newton-Raphson or the arc-length method. At the end of every converged increment we print an image of the damage contours. 
    
    \item \textbf{Damage detection}: This module is invoked at the end of every converged loading step. It receives the damage contour image and performs a series of image-analysis operations to check whether damage is present. If damage is not detected, the framework returns to the FEM solver, and the analysis continues with the next load increment. If damage is detected, this module outputs an image with the boundaries of the detected contours.
    
    \item \textbf{Damage tracking}: This module receives an image with the boundaries of the detected contours, and performs a series of operations to identify the location of the contours and scale (offset) them appropriately. This process yields a new scaled damage contour (SDC), and the output of this module is a list of all the nodal coordinates enclosed by the scaled damage contour. 
     
    \item \textbf{Domain decomposition}: This module receives the list of nodal coordinates that lie within the SDC, and then performs a series of operations to split the domain into two subdomains: a healthy group and an unhealthy group. The healthy subdomain contains only undamaged elements, $d = 0$. The unhealthy subdomain comprises of all elements which have at least one node in the aforementioned list, and therefore contains both undamaged and damaged elements, $d\geq 0$. The list of nodes that separate the healthy and unhealthy subdomains is henceforth termed \textit{interface}, and all subdomains are connected using the interior penalty method. Ultimately, the output of this module is the list of elements that constitute the healthy and unhealthy groups, which is then used to continue with the FEM analysis of the next load increment.

\end{enumerate}

Before we proceed with the detailed explanation of each module, we make a few remarks on the modularity of the overall framework. First, we note that the framework is not restricted to identifying just a single instance of damage contour. Instead, it is capable of simultaneously processing multiple damage contours that may lie apart from each other, resulting in decomposing the domain into one healthy and several unhealthy groups. This flexibility will be clearly demonstrated in the numerical examples presented in Section \ref{Sec:Numerical_Results}. Also, even though the framework currently utilizes the image-analysis techniques to aid the solution of CDM problems, this objective can be easily extended to other problems as well. For example, instead of damage contours one could use contours of von Mises plasticity, phase-field, permeability, or other state variables, expanding the scope into a very wide range of computational mechanics problems. This feature stems from the fact that the underlying mechanics problem is indifferent to the utilized image-analysis algorithms. Finally, we note that the image-printing step is optional and it is an artifact of our implementation. This could be easily replaced by a pixel-like representation of the field variable, which would still render feasible the communication between the FEM solver and the image-processing tools. 


\section{Finite Element implementation}
\label{Sec:FEM_Implementation}

To formulate the numerical solution of the governing problem with FEM, we begin by deriving the weak form of the initial problem by multiplying the strong form of Eqns. \ref{Eqn_equilibrium_1}-\ref{Eqn_equilibrium_2} with arbitrary test functions ${\boldsymbol{w}}_{u}$ and integrating over the domain. Following integration by parts and the divergence theorem we arrive at the following expression of the residual:

\begin{equation}
    {\bf{R}} = \int_{\Omega} \nabla{\boldsymbol{w}}^{T}_{u} (1 - d) \; {\mathbb{C}} \; : {\boldsymbol{\varepsilon}} \; d\Omega - \int_{\Gamma_{N}} {\boldsymbol{w}}^{T}_{u} {\boldsymbol{t}}_N d\Gamma_{N} 
\label{Eqn_residual_FEM}
\end{equation}

Using Galerkin's approximation method, the values of the state variables at the element integration points are computed as follows:

\begin{equation}
	{\bf{u}}                   = {\boldsymbol{N}}_{u} {\bf{\hat{u}}}; ~~~ 
    {\bf{\varepsilon}}         = {\boldsymbol{B}}_{u} {\bf{\hat{u}}}; ~~~
	{\boldsymbol{w}}_{u}       = {\boldsymbol{N}}_{u} {\boldsymbol{\hat{w}}}_{u}; ~~~ 
    \nabla{\boldsymbol{w}}_{u} = {\boldsymbol{B}}_{u} {\boldsymbol{\hat{w}}}_{u}
\label{Eqn_shape_functions}
\end{equation}

\noindent where ${\bf{u}}$ is the displacement field, ${\boldsymbol{N}}_{u}$ and ${\boldsymbol{B}}_{u}$ denote the displacement shape function matrix and its derivatives, and the ($\hat{.}$) symbol indicates nodal values. Combining Eqns. \ref{Eqn_residual_FEM} and \ref{Eqn_shape_functions} results to the following residual form:

\begin{equation}
    {\bf{R}} = \int_{\Omega} {\boldsymbol{B}}^{T}_{u} (1 - d) \; {\mathbb{C}} \; {\boldsymbol{B}}_{u} {\bf{\hat{u}}} \; d\Omega - \int_{\Gamma_{N}} {\boldsymbol{N}}_{u} \hat{t}_i d\Gamma_{N} 
\label{FEM_WeakRexpand2}    
\end{equation}

In order to satisfy the equilibrium condition and minimize the residual of the system, Eqn. \ref{FEM_WeakRexpand2} needs to be discretized in time and solved with a non-linear iterative scheme. In this work we employ the Newton-Raphson method and the linearized system of equations reads:

\begin{equation}
    {\bf{J}}^{i} \delta{\bf{\hat{u}}} = - {\bf{R}}^{i}
\label{Eqn_NR_system}
\end{equation}

\noindent where $i$ denotes the iteration number and $\delta{\bf{\hat{u}}}$ is the incremental change in the nodal displacement vector. The term ${\bf{J}}$ is the system Jacobian matrix, and it is generally given through a first-order Taylor approximation series as:

\begin{equation}
    {\bf{J}} = \frac{\partial{\bf{R}}}{\partial{\bf{\hat{u}}}} = {\bf{K}} + \frac{\partial{\bf{K}}}{\partial{\bf{\hat{u}}}}
\label{Eqn_Jacobian}
\end{equation}

\noindent where ${\bf{K}}$ is the secant stiffness matrix of the system, defined as:

\begin{equation}
    {\bf{K}} = \int_{\Omega} {\boldsymbol{B}}^{T}_{u} (1 - d) \; {\mathbb{C}} \; {\boldsymbol{B}}_{u} {\bf{\hat{u}}} \; d\Omega
\label{Eqn_Stiffness}
\end{equation}

In this work, due to the well-documented complications of deriving the ${\partial{\bf{K}}}/{\partial{\bf{\hat{u}}}}$ term for the non-local integral damage model \cite{jirasek2002consistent, chen2022dynamic}, we use the modified Newton-Raphson scheme throughout all the numerical examples. The Jacobian matrix is therefore approximated with the secant stiffness matrix, such as ${\bf{J}} \approx {\bf{K}}$. Using these definitions, Eqn. \ref{Eqn_NR_system} is then solved iteratively until the following convergence criterion is satisfied at each load increment: $\| \delta{\bf{\hat{u}}} \|_{2}  < 10^{-5}$, implying that the norm of the incremental changes in the displacement field vector is reduced by five orders of magnitude to achieve convergence. 

As the FEM analysis proceeds and damage propagates inside the domain, several finite elements enter their strain-softening regime. This will result into $det({\bf{J}}) \rightarrow 0$, and consequently the classical Newton-Raphson method will face challenges to converge. There is a plethora of approaches to overcome this well-established numerical issue, and without being exhaustive we mention artificial stabilization methods \cite{reese2005physically} and arc-length schemes \cite{saji2024ual}. In this work, we introduce an artificial "damping" parameter $\lambda$ as a multiplication factor of the diagonal entries of $\bf{J}$, to overcome numerical instabilities. This approach is similar in spirit with the well-known Levenberg-Marquardt Algorithm (LMA) \cite{gavin2019levenberg, roweis1996levenberg}, which introduces a damping factor for the Hessian matrix of the system. In this case, the final system of equations becomes: 

\begin{equation}
    \left({\bf{K}} + \lambda \, diag \left[ {\bf{K}} \right] \right) \delta{\bf{\hat{u}}} = - {\bf{R}}
\label{Eqn_LMA_update}
\end{equation}

At the beginning of the analysis $\lambda$ is set to 0, and the classical modified Newton-Raphson is employed. If the load incrementation step drops below $10^{-4}$, then $\lambda$ is activated and set to $10^{-9}$. If Newton-Raphson fails to converge 5 consecutive times, $\lambda$ is increased by a factor of 10. $\lambda$ is also adjusted within each increment loop, either increasing or decreasing by a factor of 10 depending on the $\| \delta{\bf{\hat{u}}} \|_{2}$ trend; more details on our numerical implementation can be found on the accompanying open-source code. In any case, $\lambda$ is capped at all times at a maximum value of $10^{-3}$. Evidently $\lambda$ remains substantially low, and its sole purpose is to allow for some numerical flexibility around the critical points in the equilibrium path, ensuring that the analysis will not be prematurely terminated. We emphasize that in our numerical examples we use identical settings between the benchmark FEM solver and the FEM solver of our framework. This ensures that the same FEM solver is used with both approaches, and any differences in the accuracy or efficiency between the two methods are safely attributed only to the image-based domain decomposition component of the proposed framework. 



\section{Damage detection}
\label{Sec:Damage_Detection}

At the end of every converged load increment we print an image of the damage contours. This image is then analyzed to identify the boundaries of the damage contour. The workflow of the damage detection module is provided in Algorithm \ref{Alg:Algorithm_damage_detection}. In the following subsections we describe all the individual processes with the aid of Fig. \ref{Figure_contour_detection_steps_elastic} (for an elastic increment, without damage) and Fig. \ref{Figure_contour_detection_steps} (for an inelastic increment, with damage). For our implementation we use the OpenCV library (which will be referred to in the code sections as cv2) \cite{opencv_library}, and we also include representative code snippets to explain each step.

\subsection{Gray-scale conversion}
\label{Sec:Gray_scale_conversion}

\begin{figure}[H]
    \centering
    \includegraphics[width=1\textwidth]{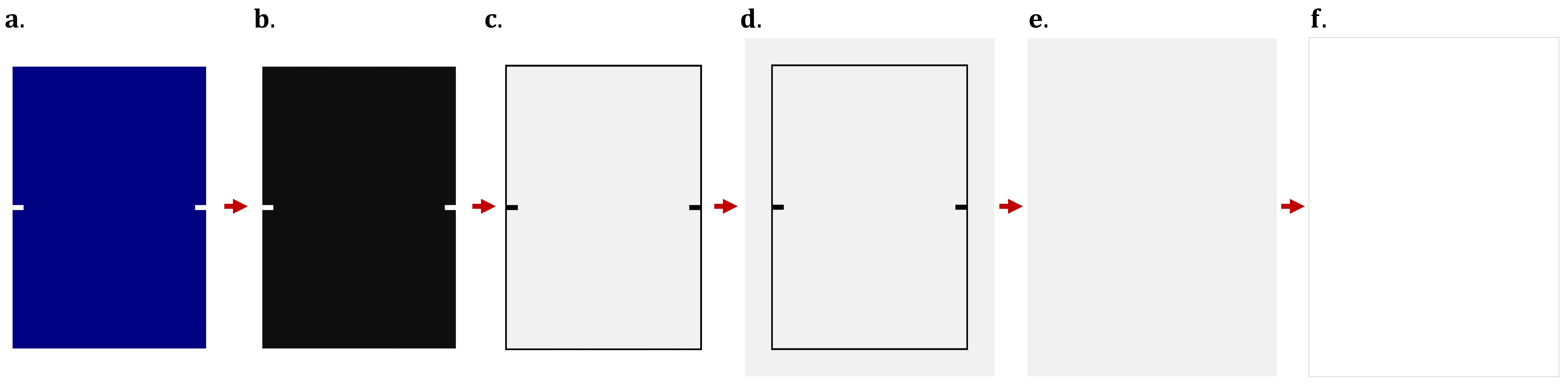}
    \caption{Contour detection steps for an elastic increment (no damage). {\bf{a.}} Input image. {\bf{b.}} After standard gray-scale conversion. {\bf{c.}} After inverting the gray-scale colormap. {\bf{d.}} After border padding. {\bf{e.}} After noise removal (cracks and boundary of initial image). {\bf{f.}} After thresholding. The image in ({\bf{f.}}) contains only white pixels, and an artificial border is added just for visual clarity.}
    \label{Figure_contour_detection_steps_elastic}
\end{figure}

\begin{figure}[H]
    \centering
    \includegraphics[width=1\textwidth]{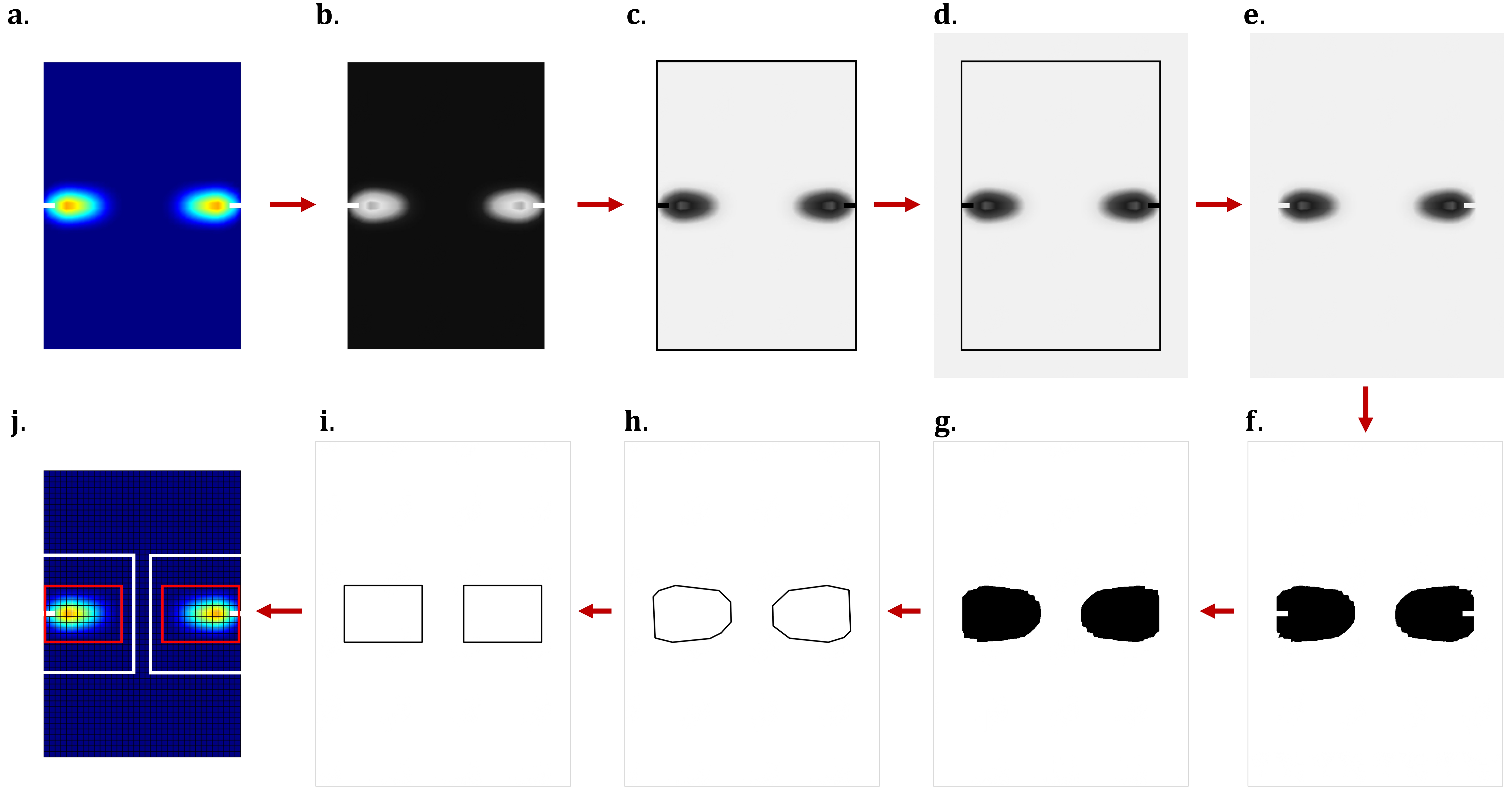}
    \caption{Contour detection steps for an inelastic increment (with damage). {\bf{a.}} Input image. {\bf{b.}} After standard gray-scale conversion. {\bf{c.}} After inverting the gray-scale colormap. {\bf{d.}} After border padding. {\bf{e.}} After noise removal (cracks and boundary of initial image). {\bf{f.}} After thresholding. {\bf{g.}} After morphological opening. {\bf{h.}} After contour detection. {\bf{i.}} After converting the detected contours into rectangles. {\bf{j}} Output image showing the contours (red rectangles) and interface nodes (white markers). The gray border in images ({\bf{f}}) - ({\bf{i}}) is added just for visual clarity.}
    \label{Figure_contour_detection_steps}
\end{figure}

The first step of the damage detection module is to load the damage contour image and convert it to a \textit{gray-scale} colormap. This step is executed as follows:

\begin{lstlisting}
import cv2
# Load the image
img = cv2.imread(path, cv2.IMREAD_GRAYSCALE)
\end{lstlisting}

\noindent where the variable $img$ is a vector representation of the image with values ranging between 0 (black) and 255 (white), and the variable $path$ is the directory of the printed image. This step has a twofold objective. First, it renders our framework independent of the colormap in which the original images are saved. This increases the flexibility of our approach and allows processing of virtually any image required. Second, gray-scale conversion leads to a reduction in the size of the data that must be stored and processed, which enhances the memory efficiency of the algorithm. The output of this step is shown in Figs. \ref{Figure_contour_detection_steps_elastic}b and \ref{Figure_contour_detection_steps}b, which show that the location and characteristics of the damage region remain unaffected as we transition from the RGB to the gray-scale representation. To enhance visual clarity we then reverse the gray-scale map as follows:

\begin{lstlisting}
# Reverse the colormap
h, w = img.shape[:2]
img = abs(255 * np.ones([h,w]) - img)
\end{lstlisting}

The outcome of this step is shown in Fig. \ref{Figure_contour_detection_steps_elastic}c and \ref{Figure_contour_detection_steps}c, where damage is shown with darker color than the rest of the domain. 

\begin{algorithm}
	\caption{Detection of damage contours}
	\label{Alg:Algorithm_damage_detection}
	\SetKwInOut{Input}{Input}\SetKwInOut{Output}{Output}
	\Input{Damage contour image}
	\Output{Contours detected}{
    Convert damage contour image to gray-scale (Section \ref{Sec:Gray_scale_conversion}) \\ 
    Pad the image to enhance visual clarity (Section \ref{Sec:Border_padding}) \\
    Remove the impact of misleading pixels (Section \ref{Sec:Seed_crack_removal}) \\
    Apply binary thresholding (Section \ref{Sec:Binary_Threshold}) \\
    Apply morphological opening (Section \ref{Sec:Morphological_Opening}) \\
    Detect the contours and transform to rectangles (Section \ref{Sec:Contour_Detection}) \\
    \uIf{no damage is found}{Proceed to the next load increment}
    \Else{Proceed to the damage tracking routine (Algorithm \ref{Alg:Algorithm_damage_tracking})}}
\end{algorithm}

\subsection{Border padding}
\label{Sec:Border_padding}

Upon a close look on Fig. \ref{Figure_contour_detection_steps_elastic}c and \ref{Figure_contour_detection_steps}c, we notice the presence of a thin dark boundary at all sides of the image. The presence of these darker pixels is an artifact of the printing software (MATLAB), resolution level and implementation process. To enhance its visualization we pad the image by adding pixels on all the image sides, as shown in Fig. \ref{Figure_contour_detection_steps_elastic}d and \ref{Figure_contour_detection_steps}d. The color of the new pixels is the \textit{median} color value of the existing pixels, and it is denoted as $t_{MCV}$. The choice of using $t_{MCV}$ for the new pixels is predicated on the logical assumption that damage is present only in a small portion of the domain and the majority of the pixels values represent no damage; thus, the median value also represents an undamaged element. This step is coded as follows:

\begin{lstlisting}
# Pad the image 
pad_size = 100
img = cv2.copyMakeBorder(img, pad_size, pad_size, pad_size, pad_size, cv2.BORDER_CONSTANT, value=[np.median(img),np.median(img),np.median(img)])
\end{lstlisting} 

\noindent where \textit{pad\_size} is a variable defining the number of added pixels on all sides of the image.

\subsection{Visual removal of misleading pixels}
\label{Sec:Seed_crack_removal}

Next, we need to eliminate the impact of all the pixels which may wrongfully be classified as representing damage. We will henceforth refer to these as \textit{misleading pixels}, and in our implementation they stem from two sources: a) the pixels constituting the thin dark 'boundary' as discussed above, which is a consequence of the printing resolution and software, and b) the pixels at the seed crack locations, which is an artifact of the model geometry. The misleading pixels can be clearly seen in Fig. \ref{Figure_contour_detection_steps_elastic}d, which is an image without damage and all pixels should have the same color, but they are also present in Fig. \ref{Figure_contour_detection_steps}d. If we do not deactivate the misleading pixels, then their darker color will prompt our contour detection algorithm to treat them as containing damage and this will clearly violate the correct damage detection.   

A simple yet effective way to eliminate their impact is the following. In the first linear elastic step of the FEM analysis, where no damage is present yet, we identify and store the \textit{coordinates} of these pixels. This step is done by comparing the color value of all the image pixels with the $t_{MCV}$, and any pixel with a color value less than $t_{MCV}$ is assigned to this group. Since these pixels are irrelevant to damage propagation, their coordinates remain constant during the analysis. Therefore, once their positions have been identified, at every subsequent load increment we directly overwrite their color values with $t_{MCV}$, imposing a no-damage condition which we know a-priori that holds true for these pixels. This step is executed as:

\begin{lstlisting}
# Deactivate the impact of "misleading" pixels
if elastic_inc == 1:
    coordinates_redundant_border = np.where(img < [np.median(img)])
    elastic_inc = 0
else:
    coordinates_redundant_border = coordinates_redundant_border_stored
img[coordinates_redundant_border] = np.median(img)
\end{lstlisting}

\noindent where the variable \textit{elastic\_inc} is a flag indicating if this load increment is the first one, thus belonging to the elastic zone. Fig. \ref{Figure_contour_detection_steps_elastic}e and \ref{Figure_contour_detection_steps}e show the image after the misleading pixels have been removed. We observe that we have now successfully isolated the actual damage regions.

\subsection{Binary thresholding}
\label{Sec:Binary_Threshold}

The next step in the damage detection module is to apply binary thresholding, one of the most fundamental image-segmentation techniques \cite{sezgin2004survey}. Essentially, this step turns our gray-scale image into a black-and-white one, with black indicating all pixels where damage is detected and white reflecting the opposite. This is achieved by satisfying the following condition: if the pixel color value is less than a threshold $t_{TH}$ then its color value is replaced by 0 (black - background), otherwise it is replaced by 1 (white - foreground). The threshold value $t_{TH}$ is a user-defined hyperparameter, and in our implementation we have set $t_{TH} = t_{MCV} - 1$. This evidently imposes a very strict criterion, classifying as unhealthy all the pixels whose color is less than median PCV minus one, and therefore it can capture even the slightest changes in the damage contours. Fig. \ref{Figure_contour_detection_steps_elastic}f and \ref{Figure_contour_detection_steps}f show the resulting images after binary thresholding has been applied, and this step is executed as:

\begin{lstlisting}
# Apply binary thresholding
_, img = cv2.threshold(img, np.median(img)-1, 255, cv2.THRESH_BINARY)
\end{lstlisting}

\subsection{Morphological opening}
\label{Sec:Morphological_Opening}

As we see in Fig. \ref{Figure_contour_detection_steps}f, it is possible after thresholding that some small white pixels may be present inside the black pixel region. These may represent the seed cracks, or they may stem from a less strict $t_{TH}$ criterion. Since their presence may introduce undesired consequences in the subsequent contour detection step, we remove them by applying \textit{morphological opening} \cite{vincent1994morphological}, a widely used image-convolution operation. During image convolution, the color value of each pixel in the original image is replaced by the sum of an element-wise multiplication between the kernel and the color values of the neighbouring pixels in the image. The convolution operation is given as \cite{jeon2017active}: 

\begin{equation}
    f(x,y) = g * o(x,y) = \sum_{i=-a}^{a} \sum_{j=-b}^{b} \; g(i,j) \; o(x-i,y-j) 
\end{equation}

\noindent where $f(x,y)$ is the filtered image, $o(x,y)$ is the original image, $g$ is the kernel, $(2a + 1, 2b + 1)$ are the dimensions of the kernel, $(x,y)$ denote the row and column indices in the original and filtered images, $(i,j)$ are the row and column indices in the kernel matrix, and $(*)$ denotes the element-wise multiplication. In our implementation, we assume a square kernel $g$ of ones with size $g_{size} = 2a + 1 = 2b + 1$. 

Morphological opening involves two steps: an erosion followed by a dilation. In erosion, the kernel slides over the image and overwrites the values of the pixels as follows: the new color is 1 (white) if all the pixels under the kernel have a value of 1, otherwise it is set to 0 (black). In the subsequent dilation process, the kernel slides over the image and the new pixel color is 1 if any pixel under the kernel is 1, otherwise its value is set to 0. Ultimately, this process removes the white pixels inside the detected damage regions as shown in Fig. \ref{Figure_contour_detection_steps}g. 

\begin{lstlisting}
# Apply morphological opening
kernel_size = 11
num_iterations  = 5
kernel = cv2.getStructuringElement(cv2.MORPH_RECT,(kernel_size,kernel_size))
img = cv2.morphologyEx(img, cv2.MORPH_OPEN, kernel, iterations=num_iterations)
\end{lstlisting}

\subsection{Contour detection, smoothing and rectangular reshaping}
\label{Sec:Contour_Detection}

The next step of the damage detection module is to apply a contour detection algorithm and identify the boundaries of the damage region. Here we use the contour detection method of OpenCV \cite{opencv_library}, which relies on the border following algorithm proposed by \cite{suzuki1985topological}. The latter algorithm was developed for binary images, and therefore it is particularly suitable for our black-and-white samples. The outcome of this step is the coordinate list of all the pixels that constitute the damage contour. At this stage, depending on the printing resolution and the user-defined hyperparameters, the shape of the detected contour may be highly segmented. We can then approximate the detected contour with a smoother polygon that comprises of less vertices, in order to facilitate the subsequent processes of rectangular transformation and offset-scaling. The OpenCV method for this step uses the Ramer–Douglas–Peucker algorithm, and Fig. \ref{Figure_contour_detection_steps}h shows the detected contours at this stage. The code snippet for detecting and smoothing the contours reads as:  

\begin{lstlisting}
# Identify contours
contours, _ = cv2.findContours(img, cv2.RETR_TREE, cv2.APPROX_SIMPLE)
# Approximate with smoother polynomials
contours = cv2.approxPolyDP(cnt, 0.01*cv2.arcLength(cnt, True), True) for cnt in contours
\end{lstlisting}

Finally, we transform the detected contours into ones that have a rectangular shape. The reason behind this modification is that in the subsequent damage-tracking module we need to offset-scale the contours, and this process is executed in a more straightforward fashion using rectangular shapes. Fig. \ref{Figure_contour_detection_steps}i depicts the final output of the damage detection module, showing the rectangular damage contours, and this step is coded as:

\begin{lstlisting}
# Transform into rectangular contours
for i in range(len(contours)): 
        min_dir_x = min((contours[i])[:, 0, 0]); max_dir_x = max((contours[i])[:, 0, 0])
        min_dir_y = min((contours[i])[:, 0, 1]); max_dir_y = max((contours[i])[:, 0, 1])
        contours[i] = np.asarray([[[min_dir_x, min_dir_y]], [[max_dir_x, min_dir_y]], 
                                    [[max_dir_x, max_dir_y]], [[min_dir_x, max_dir_y]]])
\end{lstlisting}

Finally, we note here that another reason behind introducing the modification steps after our damage contour detection is to enhance the framework modularity and thus increase the user-flexibility. Depending on the problem, the user may need to include or omit the smoothing and rectangle-reshaping steps, and therefore we choose to provide additional functionalities in our framework. 

\section{Damage tracking}
\label{Sec:Damage_Tracking}

The purpose of the \textit{damage tracking} module is to process the information of the detected contours and decide whether a new domain decomposition is warranted. A step-by-step execution of this sub-routine is presented in Algorithm \ref{Alg:Algorithm_damage_tracking}. To aid the readability of this section, we introduce the following nomenclature:

\begin{itemize}
    \item DC - damage contour obtained from the image detection module (Section \ref{Sec:Damage_Detection})
    \item SDC - scaled damage contour obtained after appropriate scaling of the DC
    \item PSDC - scaled damage contour from the previous load increment
\end{itemize}

All contours have a rectangular shape, and $xc_{DC}$ and $yc_{DC}$ are the lists of x- and y- coordinates of the DC corners. 

\subsection{Damage directionality and scaling}
\label{Sec:Damage_directionality}

\begin{figure}[H]
    \centering
    \includegraphics[width=1\textwidth]{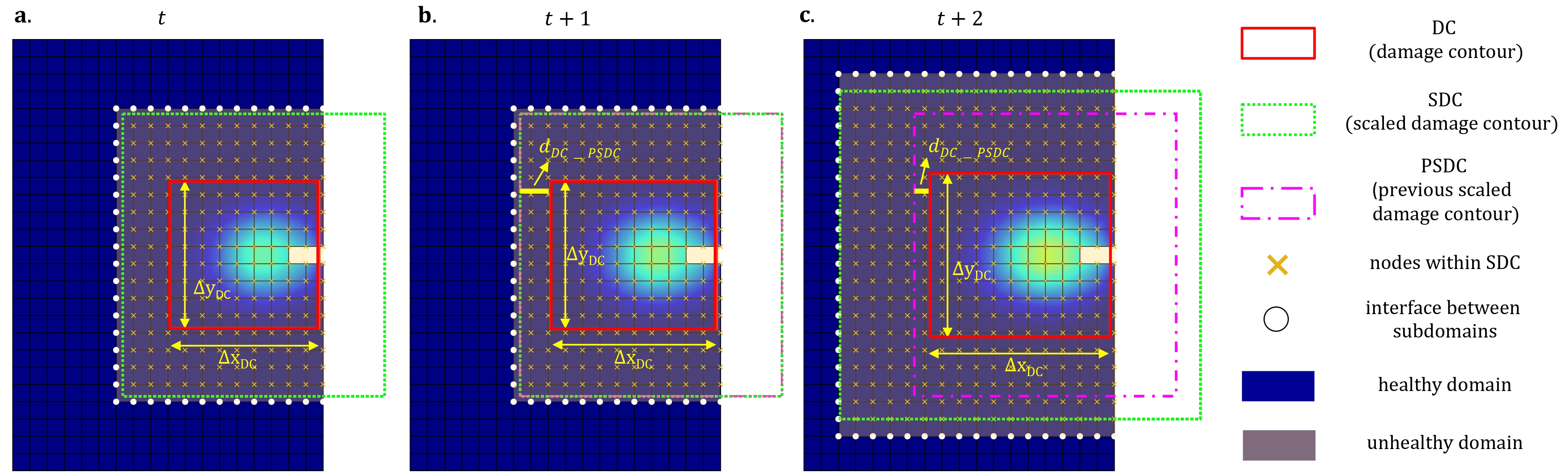}
    \caption{{\bf{a.}} Schematic view of DC (red), SDC (green) and interface nodes (white markers) at load increment $t$. {\bf{b.}} Similar schematic for the next load increment $t+1$. The magenta curve is PSDC, the scaled contour from (a). Here $d_{DC-PSDC} > d_{thres}$, therefore the new scaled contour (green) coincides with the PSDC (magenta). {\bf{c.}} Similar schematic for the next  load increment $t+2$. Here $d_{DC-PSDC} < d_{thres}$, therefore a new scaling and domain decomposition is warranted.}
    \label{Figure_algorithm_damage}
\end{figure}

First, we need to identify the directionality of the damage path since this is a prerequisite for the optimum offset scaling of its contour. We use the information from the damage contour shape and we compute the DC dimensions in the $x$ and $y$ directions (indicated in Fig. \ref{Figure_algorithm_damage}) as follows:

\begin{equation}
    \Delta x_{DC} = max(xc_{DC}) - min(xc_{DC}) \; \; \; ; \; \; \; 
    \Delta y_{DC} = max(yc_{DC}) - min(yc_{DC})
\label{Eqn_DC_dimensions}
\end{equation}

Next, we scale the DC appropriately in order to ensure that it is surrounded by an optimal number of healthy elements. The number of healthy elements around the damaged region plays a vital role in the efficiency of our framework. On one side, selecting a large number of undamaged elements could degrade the efficiency of the algorithm by increasing the computational cost. On the contrary, choosing very few elements could impact the quality of the numerical solution, since damage would then be present at the interface between the healthy and unhealthy zones. Let us define the vector $sf$, which contains scaling factors in the $x$ and $y$ directions respectively: $sf = $ [$sf_{x} \: ; \: sf_{y}$]. We also define the user-defined scalar values $sf_{user}$ and $sf_{thresh}$. The variable $sf_{user}$ represents the scaling magnitude in the direction where damage propagates more, and the variable $sf_{thresh}$ acts as the minimum scaling factor in the other direction. We then distinguish between three cases:

\begin{algorithm}[t!]
	\caption{Damage Tracking}
	\label{Alg:Algorithm_damage_tracking}
    \If{no new contours are detected in this step}
    {\uIf{damage has not been detected overall}
         {Set: decomposition flag = False; \bf{exit}}
         \Else{
         \For{each contour detected}{
            Sort contours based on their centroid \\
            Compute $d_{DC-PSDC}$: minimum distance between current damage contour (DC) and scaled damage contour from the previous increment (PSDC) \\
            \uIf{$d_{DC-PSDC} > d_{thres}$}{
            Set: decomposition flag = False; \bf{exit}}
            \Else{\If{$d_{DC-PSDC} = 0$}{Set: repeat\_increment flag = True; \bf{break}}}
            }}}
    \If{at least a new contour is detected in this step, \normalfont{{\bf{or}}} $0 < d_{DC-PSDC} \leq d_{thres}$}{
    \For{each contour}{Compute $\Delta x_{DC}$, $\Delta y_{DC}$ (Eqn. \ref{Eqn_DC_dimensions}) \\
    Scale the contours accordingly (Eqn. \ref{Eqn_scaling_factors}) \\
    Check if the new scaled contours are intersecting, if yes then merge \\
    Obtain the FEM nodes inside the scaled contours \\
    Set: decomposition flag = \bf{True}
    }}
\end{algorithm}

\begin{equation}
    sf = \begin{cases}
        \left[sf_{user} \; \; ; \; \; max(\left(\Delta y_{DC} / \Delta x_{DC}\right) * sf_{user}, sf_{thresh})\right] & if \; \; \Delta x_{DC} > \Delta y_{DC} \\
        \left[max(\left(\Delta x_{DC} / \Delta y_{DC} \right) * sf_{user}, sf_{thresh}) \; \; ; \; \; sf_{user}\right] & if \; \; \Delta x_{DC} < \Delta y_{DC} \\
        \left[sf_{user} \; \; ; \; \; sf_{user}\right] & if \; \; \Delta x_{DC} = \Delta y_{DC}
\end{cases}
\label{Eqn_scaling_factors}
\end{equation}

Evidently, Eqn. \ref{Eqn_scaling_factors} assigns different scaling factors in the $x$ and $y$ directions depending on the direction in which damage exhibits more growth. This approach provides sufficient flexibility to the end-user on tuning the scaling parameters depending on the investigated problem. The DC dimensions are then multiplied with the scaling factors from Eqn. \ref{Eqn_scaling_factors} as: $\Delta x_{SDC} = \Delta x_{DC} \times sf_{x}; ~\Delta y_{SDC} = \Delta y_{DC} \times sf_{y}$, arriving at the scaled damage contours SDC denoted with the green dashed lines in Fig. \ref{Figure_algorithm_damage}.

\subsection{When to decompose?}
\label{Sec:When_to_decompose}

An important aspect of damage tracking involves deciding when a new decomposition is needed, since not every converged load step warrants a new decomposition. A new decomposition is required only when: 

\begin{itemize}
    \item A damage contour is detected for the first time, or 
    \item The minimum distance between the DC and the scaled damage contour of the previous increment (PSDC) is below a user-specified distance: $d_{DC-PSDC} < d_{thres}$.
\end{itemize}

\begin{figure}[H]
    \centering
    \includegraphics[width=0.8\textwidth]{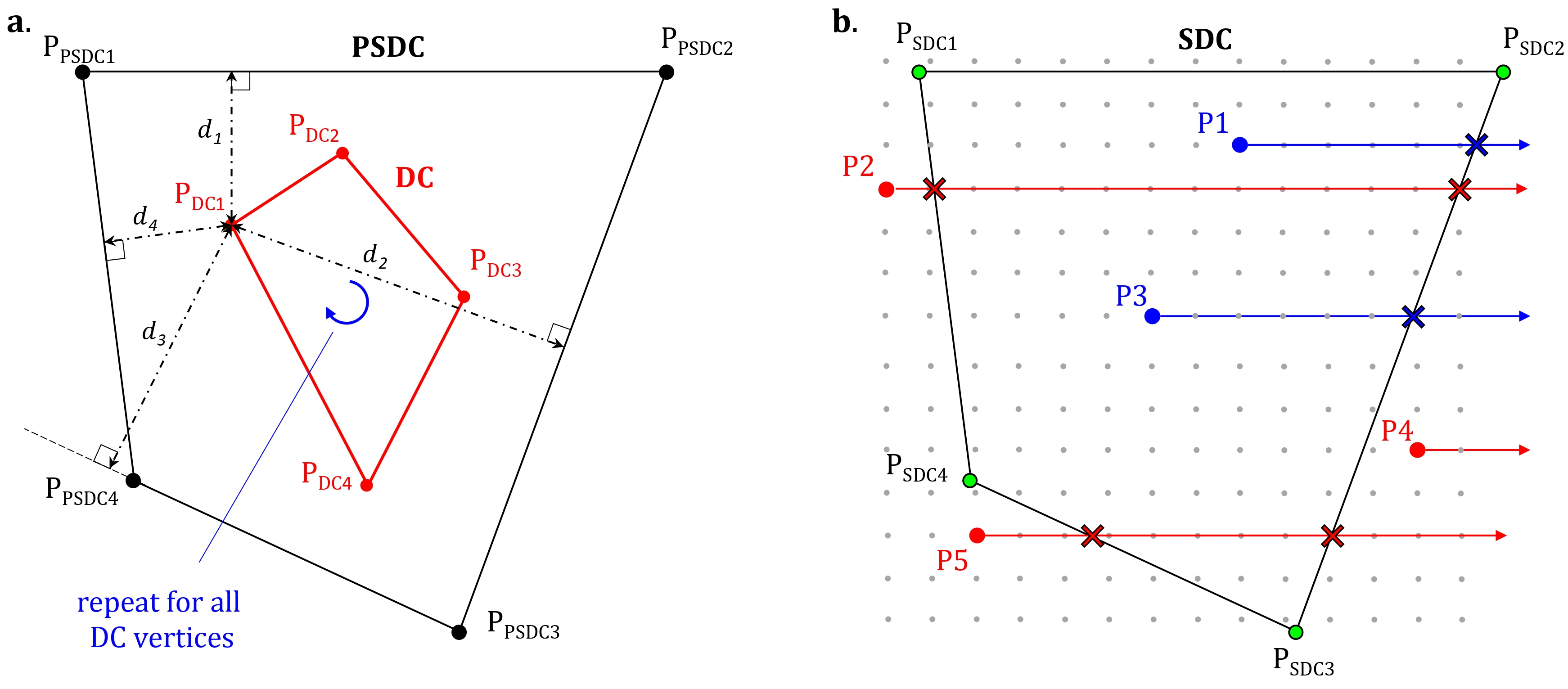}
    \caption{{\bf{a.}} Calculating the minimum distance between the damage contour (DC, red polygon) and the scaled damage contour from the previous increment (PSDC, black polygon). For all vertices in DC we compute the distances to the line segments of PSDC. We then repeat the same process for all DC vertices and PSDC edges. The minimum of all these distances is the queried $d_{DC-PSDC}$. {\bf{b.}} Ray-casting algorithm. For each node of the domain (gray circles) we draw a line to the positive infinity, and count the number of intersections with the scaled damage contour (SDC) boundary. If the number of intersections is odd, then the node lies inside the SDC (blue points). If this number is even (zero included), then it lies outside (red points).}
    \label{Figure_mindist_raycasting}
\end{figure}

While justifying the first condition is evident, the objective of the second condition is to reduce the number of times that decomposition takes place, and therefore decrease the total computational expense. DC and PSDC are both polygons, and to compute the minimum distance $d_{DC-PSDC}$ between them we use a slightly modified version of the algorithm in \cite{jacquenot_mindist}. We briefly describe the original algorithm and our modification below, while referring to the schematic in Fig. \ref{Figure_mindist_raycasting}a. In that figure the DC is shown with red color (inner rectangle) and the PSDC is shown with magenta color (outer rectangle). 

First, the algorithm checks whether the polygons intersect with each other, in which case the minimum distance is considered zero. If the polygons do not intersect, then starting with the first DC vertex, $P_{DC1}$, we compute its distance to the four line segments of PSDC by finding the projection of $P_{DC1}$ on these lines. This process is repeated for every vertex point of the DC ($P_{DC1} - P_{DC4}$), as well as for all the PSDC vertices ($P_{PSDC1} - P_{PSDC4}$) against the DC lines. The minimum of all these values is the queried distance $d_{DC-PSDC}$. In our slightly modified implementation, we assign importance factors to the distances in the $x$ and $y$ directions using directional weights. This is because we want to place more emphasis on the direction of damage growth, while reducing the impact of the perpendicular direction. For this reason, we define the weight vector $w = $ [$w_{x} \: ; \: w_{y}$] as follows:

\begin{equation}
    w = \begin{cases}
        \left[1 \; \; \; ; \; \; \; \dfrac{sf_{x}}{sf_{y}}\right]   & if \; \; \Delta x_{DC} > \Delta y_{DC} \\
        \left[\dfrac{sf_{y}}{sf_{x}} \; \; \; ; \; \; \; 1\right]   & if \; \; \Delta x_{DC} < \Delta y_{DC} \\
        \left[1 \; \; \; ; \; \; \; 1 \right]  & if \; \; \Delta x_{DC} = \Delta y_{DC}
\end{cases}
\label{Eqn_weight_factors}
\end{equation}

The weights given in Eqn. \ref{Eqn_weight_factors} provide a measure of the relative growth of damage in the $x$ and $y$ directions, and they are used as multipliers of the respective distances (our implementation can be found in the accompanying code). We note that if the end-user chooses the scaling factors to be $sf_{user} = sf_{thres}$, then the directional weights become $w_{x} = w_{y} = 1$, thus neglecting the impact of this additional functionality. 

\subsection{Ray casting}
\label{Sec:Ray_casting}

The last step of the damage tracking module is to determine the nodes that are encircled by the SDC. This is a problem similar to determining whether a point lies within a polygon, and it can be solved using the \textit{ray casting} algorithm \cite{hormann2001point}. For our implementation we use the algorithm provided in \cite{ray_tracing_geeks}, which we briefly explain below. This algorithm performs the following steps: it iterates through every node of the domain, draws a horizontal line from the right side of that node to infinity, and then checks the number of intersection points between the horizontal line and the SDC boundary. If the number of intersections is odd then the node lies inside the polygon, which implies that it belongs to the damaged subdomain. If the number of intersections is even, then the queried node is outside the SDC. A schematic view of this step is shown in Fig. \ref{Figure_mindist_raycasting}b. This process is repeated for all the nodes, and the final outcome of the ray casting algorithm is the list of nodes that lie within the scaled damaged contour.

\section{Domain decomposition}
\label{Sec:Domain_Decomposition}

The last module of our framework involves the \textit{domain decomposition} component. Here, we construct the healthy and the damaged subdomains, and leverage on the Schur complement to realize the computational savings in our numerical solution. Below we elaborate on these processes.

Once the ray casting algorithm has determined the list of nodes inside the SDC, we iterate through the elements of the domain and check whether any node in their mesh connectivity belongs to that list. Any element that has at least one node in the list is assigned to the unhealthy subdomain, and the remaining are assigned to the healthy subdomain. Numerically, we satisfy the continuity requirement between the healthy and unhealthy subdomains using the penalty method, which is a widely used approach for domain decomposition \cite{bautista2015nonconformal, shi2020penalty} and contact \cite{zang2011contact, chouly2013convergence} problems. Following the implementation of the interior penalty method, artificial springs with a high stiffness value $\beta$ are used to penalize the differential response of the nodes that are shared between the domains (overlapping nodes). In our examples, $\beta$ is set to be 4 orders of magnitude greater than the maximum stiffness value of the healthy domain; more details on the implementation can be found in the accompanying code. 

After the finite elements have been assigned either to the damaged or the undamaged subdomains, the global system of equations is re-arranged as follows: 

\begin{equation}
    \begin{array}{ccc}
    \begin{bmatrix}
    {\bf{K}}_{HH} & {\bf{K}}_{HU} \\
    {\bf{K}}_{UH} & {\bf{K}}_{UU} \\
    \end{bmatrix}
    &
    \begin{Bmatrix}
    \delta {\bf{u}}_H \\
    \delta {\bf{u}}_U \\
    \end{Bmatrix}
     = -
     &
    \begin{Bmatrix}
    {\bf{R}}_H \\
    {\bf{R}}_U \\
    \end{Bmatrix}
    \end{array}
\label{Eqn_global_system_two_domains}
\end{equation}

\noindent where the subscripts $H$ and $U$ refer to the healthy and unhealthy subdomains respectively. ${\bf{K}}_{HH}$ and ${\bf{K}}_{UU}$ are the stiffness matrices of the associate domains. The entries of ${\bf{K}}_{UH}$ and ${\bf{K}}_{HU}$ are either zero or $\pm \beta$, and they represent the connectivity between the healthy and the unhealthy domains. The system of Eqn. \ref{Eqn_global_system_two_domains} can be solved efficiently using the Schur complement approach \cite{zhang2006schur}, which is a widely used method for domain decomposition. Using this approach, we re-arrange Eqn. \ref{Eqn_global_system_two_domains} to arrive at the following expressions: 

\begin{subequations}

    \begin{equation}
        {\bf{S}} = {\bf{K}}_{UU} - {\bf{K}}_{UH} \left[ {\bf{K}}_{HH} \right]^{-1}  {\bf{K}}_{HU}
    \label{Eqn_Schur_S}
    \end{equation}
    
    \begin{equation}
        \delta {\bf{u}}_U = \left[{\bf{S}}\right]^{-1} \left( - {\bf{R}}_{U} + {\bf{K}}_{UH} \left[{\bf{K}}_{HH}\right]^{-1} {\bf{R}}_{H} \right) 
    \label{Eqn_Schur_duu}
    \end{equation}
    
    \begin{equation}
        \delta {\bf{u}}_H = - \left[{\bf{K}}_{HH} \right]^{-1} \left( {\bf{R}}_{H}
        + {\bf{K}}_{HU} \delta {\bf{u}}_U \right)
    \label{Eqn_Schur_duh}
    \end{equation}
    
\label{Eqn_Schur}
\end{subequations}

Here we underline that Eqns. \ref{Eqn_Schur_duu} and \ref{Eqn_Schur_duh} are solved using the \textit{mldivide} MATLAB function on ${\bf{K}}_{HH}$, and the latter operation takes place only at the first iteration of a load increment where a new decomposition is warranted. Otherwise, this quantity is stored in memory and used for the computation of $\delta {\bf{u}}_U$ and $\delta {\bf{u}}_H$, until the damage-tracking module dictates a new domain decomposition. Overall, this is a significantly faster numerical procedure than the iterative solution of Eqn. \ref{Eqn_global_system_two_domains}.

\section{Results}
\label{Sec:Numerical_Results}

We test our framework against a series of models, and in this section we present the results of our numerical investigation. Table \ref{Table_List_of_cases} shows an aggregate list of all the examples, along with their specific modeling details. We compare the results of our domain decomposition approach (DD) against the FEM analysis of the single domain idealization (SD). The following hyperparameters are common across all problems: we assume plain strain conditions, the elastic properties are G = 125 MPa (shear modulus) and $v = 0.2$ (Poisson's ratio), and the Mazars damage law is adopted (Eqn. \ref{Eqn_Mazars}) with parameters $\alpha = 0.8$ and $\beta = 20000$. Unless otherwise specified, the convergence tolerance is $tol = 10^{-5}$, the colormap for printing is $jet$, and the image detection hyperparameters are $sf_{user} = 2$, $sf_{thres} = 2$, $d_{thres} = 1mm$.

\begin{table}[H]
    \centering
    \begin{tabular}{c c c c c c c c c}
        \hline
        {\bf{Section}} & {\bf{Geometry}} & {\bf{Mesh shape}} & $\bar{u}$ & {\bf{Damage law}} & $\bf{d_{max}}$ & $\bf{l_{c}}$ & ${\bf{\varepsilon}}_{eq}$ \\ 
        \hline
        \multirow{2}{*}{\ref{Sec:SNT}} & \multirow{2}{*}{SNT}  & structured   & $\pm$ 0.0013  & local     & 0.99  & -   & Eqn. \ref{Eqn_eqstrain_tensile}  \\
         &   & unstructured & $\pm$ 0.0021  & non-local & 0.99  & 3.0 & Eqn. \ref{Eqn_eqstrain_shear}    \\
        \ref{Sec:SDNT} & SDNT & structured   & $\pm$ 0.00175 & non-local & 0.999 & 2.5 & Eqn. \ref{Eqn_eqstrain_shear}    \\
        \ref{Sec:ADNT} & ADNT & structured   & $\pm$ 0.00175 & non-local & 0.99  & 2.5 & Eqn. \ref{Eqn_eqstrain_shear}    \\
        \hline    
    \end{tabular}
    \caption{List of all the examples investigated. For each problem we use three mesh idealizations (Coarse, Intermediate, Fine), the details of which are found in the respective sub-sections. For all cases we compare the performance of our framework against the conventional SD approach.}
    \label{Table_List_of_cases}
\end{table}

\subsection{Single Notch under Tension (SNT)}
\label{Sec:SNT}

\begin{figure}[H]
    \centering
    \includegraphics[width=0.7\textwidth]{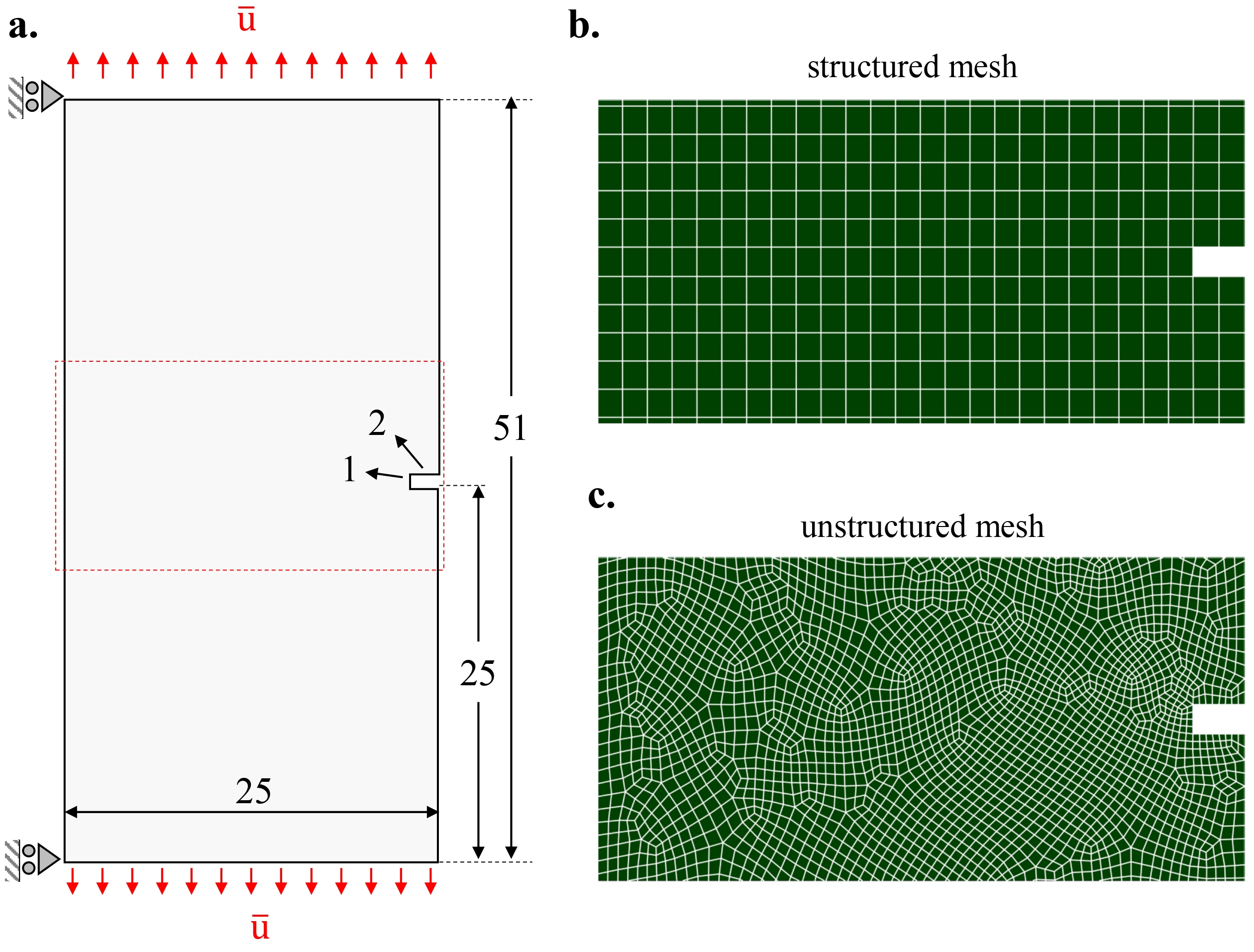}
    \caption{{\bf{a.}} Geometric and loading details of the Single Notch Tension (SNT) model. The FEM mesh of the region enclosed by the red line is shown in ({\bf{b}}) and ({\bf{c}}) for the first (structured mesh) and second (unstructured mesh) variations respectively.}
    \label{Figure_SNT}
\end{figure}

The first numerical model is a Single Notch under Tension (SNT) problem. Fig. \ref{Figure_SNT}a shows a schematic view of its geometry and loading conditions. The top and bottom faces are subject to a prescribed tensile displacement $\bar{u}$, and the top left and bottom left nodes are constrained in the horizontal direction. For the SNT problem we investigate two variations, which differ in terms of mesh shape regularity, equivalent strain definition, and damage non-locality. These variations are presented below.

\subsubsection{SNT with structured mesh and local damage}
\label{Sec:SNT_struct_local}

For the first variation of the SNT problem we adopt a structured mesh, the local damage law, and the equivalent strain definition of Eqn. \ref{Eqn_eqstrain_tensile}. This model is termed $SNT_{struct}$, it has 1376 finite elements, and a closeup view of its mesh idealization at the domain middle part is shown in Fig. \ref{Figure_SNT}b. The goal of this variation is to showcase the accuracy and computational savings of the proposed framework, as well as to examine the impact of several algorithmic hyperparameters. 

\begin{figure}[b!]
    \centering
    \includegraphics[width=1\textwidth]{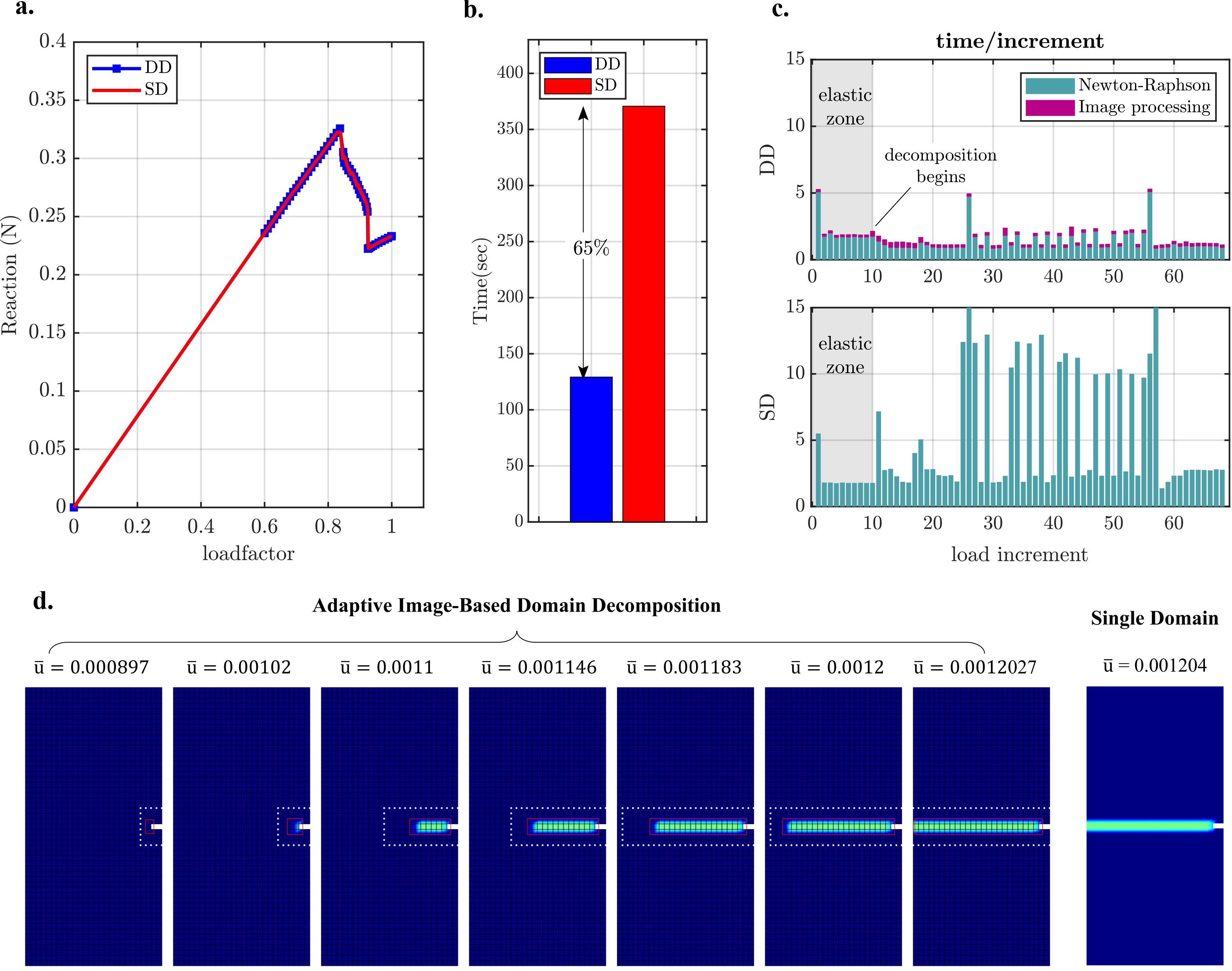}
    \caption{Numerical results of the $SNT_{struct}$ model. Comparison between our domain decomposition framework (DD) and conventional single domain (SD) analysis, showing in {\bf{a.}} reaction-loadfactor response, {\bf{b.}} total simulation times and {\bf{c.}} time/increment. {\bf{d.}} Snapshots of damage propagation using our adaptive domain decomposition framework.}
    \label{Figure_SNTstruct_results}
\end{figure}

We compare our domain decomposition (DD) approach with a conventional single domain (SD) finite element analysis, and we report the results in Fig. \ref{Figure_SNTstruct_results}. The reaction force curves are plotted in Fig. \ref{Figure_SNTstruct_results}a, and the evident overlap between the two curves provides the first evidence of our method accuracy. Additionally, in Fig. \ref{Figure_SNTstruct_results}d we present several snapshots of damage propagation captured with our method, which demonstrate in practice the adaptive domain decomposition component of our framework. For comparison, we show the damage contour of the SD analysis when the crack has fully propagated throughout the domain. Evidently, the shape and size of the damage contours generated with the two methods at that load level match with each other, and similar observations can be made after inspecting the contours at any load increment. Overall, these findings prove that the proposed method holds the same level of accuracy as the conventional solver. 

Having established confidence in the accuracy of our solution, we move to the comparison of the computational performance. In Fig. \ref{Figure_SNTstruct_results}b we plot the total time taken by the two methods, and we report computational savings of $65\%$ for this problem. A closer investigation of the simulation cost is conducted in Fig. \ref{Figure_SNTstruct_results}c, where we display the time/increment for the two methods.
The light petrol bars represent the time taken by the Newton-Raphson algorithm, while the dark magenta bars represent the cost of image detection and domain decomposition modules - these are present only in our method. Here we can make two remarks. First, we note that the latter modules are fully active from the onset of the analysis, and their contribution to the total cost is substantially less than their Newton-Raphson counterpart. This shows that our image-based tools impose a rather insignificant overhead in the analysis, and this important feature will be even more evident in the next example where we benchmark our method at finer mesh idealizations. Secondly, when we are still in the elastic zone (highlighted with gray color), we observe that the time/increment is approximately the same between the two methods. However, once damage is detected at the $10^{th}$ load increment the domain decomposition module is activated, and from that point onwards we can observe the computational savings gained at each increment.

\begin{figure}[b!]
    \centering
    \includegraphics[width=0.8\textwidth]{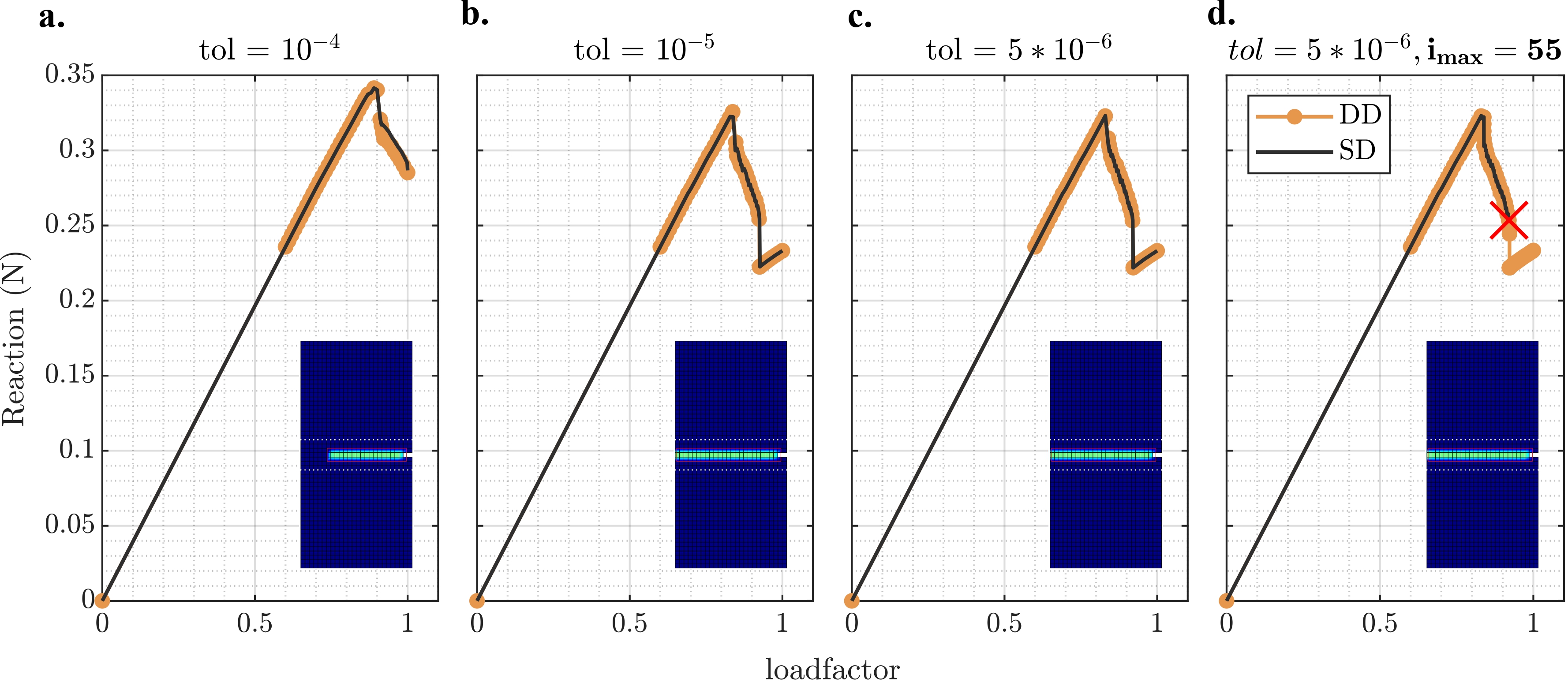}
    \caption{Reaction-loadfactor curves for different values of convergence tolerance; comparison between DD and SD. Inset plots show the damage contour at the last increment obtained with our framework. The maximum number of iterations/increment in the analyses of ({\bf{a.}}) - ({\bf{c.}}) is  $i_{max} = 150$. The red cross in ({\bf{d.}}) indicates the last converged increment of the SD analysis.}
    \label{Figure_SNT_conv_tolerance}
\end{figure}

In order to explore the robustness of our framework, as well as to establish further confidence in the choice of convergence tolerance, we launch a parametric study on the latter quantity. We explore three cases: $tol = 10^{-4}, 10^{-5}$ and $5 * 10^{-6}$. Fig. \ref{Figure_SNT_conv_tolerance}a-c shows the reaction forces for the three cases respectively, for both DD and SD, and the inset images represent the damage contours at the last converged increment of our DD analysis. Overall, the DD curves always coincide with their SD counterparts, which further verifies that our approach follows the same level of accuracy as the conventional solution. Also, we note that the most relaxed tolerance $tol = 10^{-4}$ overshoots the force peak and it's not able to capture the damage spread throughout the entire domain, whereas the other two tolerance values yield almost identical results. Therefore, this verifies the correctness of using tol = $10^{-5}$ for our simulations. Finally, we perform one additional experiment by fixing the maximum number of allowable iterations per increment to $i_{max} = 55$ (this number was $i_{max} = 150$ in the analyses of Fig. \ref{Figure_SNT_conv_tolerance}a-c). This is a stricter numerical constraint and the results of this investigation are plotted in Fig. \ref{Figure_SNT_conv_tolerance}d. Here, we observe that the SD analysis is not able to capture the vertical drop in the force diagram, failing to converge after several attempts. On the contrary, our DD approach is able to overcome this critical point and proceed further in the equilibrium path, which clearly demonstrates the robustness of the domain decomposition method over the single domain analysis.

\begin{figure}[H]
    \centering
    \includegraphics[width=1\textwidth]{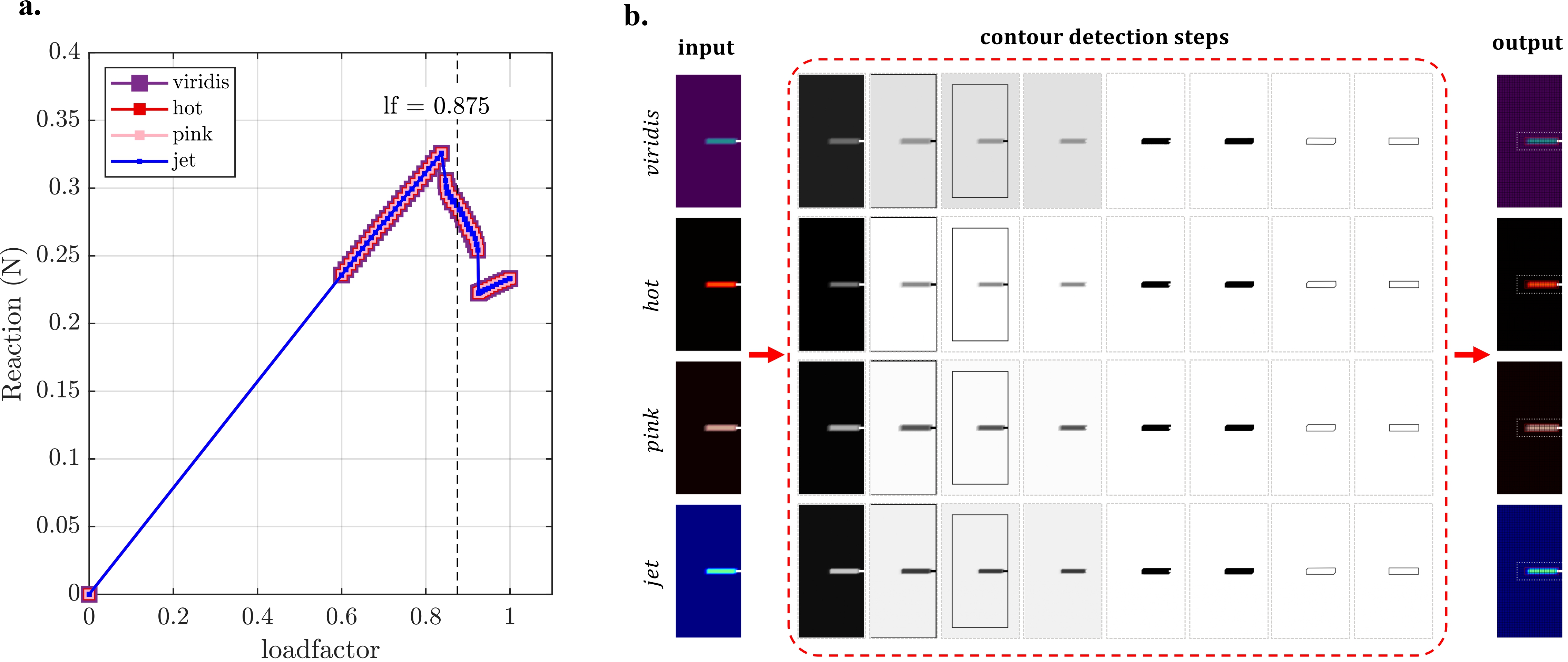}
    \caption{Evidence of colormap independence: {\bf{a.}} Reaction-loadfactor curves for the first SNT problem using four different colormaps. The coinciding curves show that our framework is insensitive to the choice of colormap. {\bf{b.}} Snapshots of the damage contour detection algorithm steps for the four colormaps. The last black-and-white images that contain the detected contour are identical across all cases.}
    \label{Figure_SNT_colormap}
\end{figure}

At this stage, it is also crucial to ensure that our framework is not sensitive to the choice of the colormap used for image printing. Establishing this functionality proves that our method is not constrained by this user-specific choice, a feature which broadens significantly the scope of candidate images that can be analyzed with our approach. To this end, we launch a parametric study on four representative colormaps available in MATLAB: colormap = $viridis$, $hot$, $pink$ and $jet$. Fig. \ref{Figure_SNT_colormap}a shows the reaction forces for these cases, and we observe that they all coincide exactly with each other. To further demonstrate the colormap insensitivity, we present in Fig. \ref{Figure_SNT_colormap}b the image processing steps at a representative load increment across the four cases. The sequence of images matches that of Fig. \ref{Figure_contour_detection_steps}, and the reader is referred to that figure for the explanation of each step's functionality. We observe that the gray-scale conversion of each colormap produces a different image, as evidenced by the different shades of gray in steps 1-4. However, the binary thresholding step (along with our choice of $t_{TH}$) enables the identification of all the pixels that represent damage; hence the identical images in steps 5-8 for all colormaps. Consequently, in all cases the domain decomposition module receives indistinguishable input, justifying our framework's insensitivity to the colormap choice.

\subsubsection{SNT with unstructured mesh and non-local damage}
\label{Sec:SNT_unstruct_nonlocal}

In the second variation of the SNT problem we use an unstructured mesh, the non-local integral damage law, and the equivalent strain definition of Eqn. \ref{Eqn_eqstrain_shear}. This model is termed $SNT_{unstruct}$, it has 8588 finite elements, and Fig. \ref{Figure_SNT}c shows a closeup view of the mesh idealization at the domain middle part. The purpose of this variation is to demonstrate that our framework can be readily applied to cases with an irregular mesh discretization, which drastically expands the spectrum of candidate problems, as well as that it is compatible with non-local damage representations. 

\begin{figure}[b!]
    \centering
    \includegraphics[width=1\textwidth]{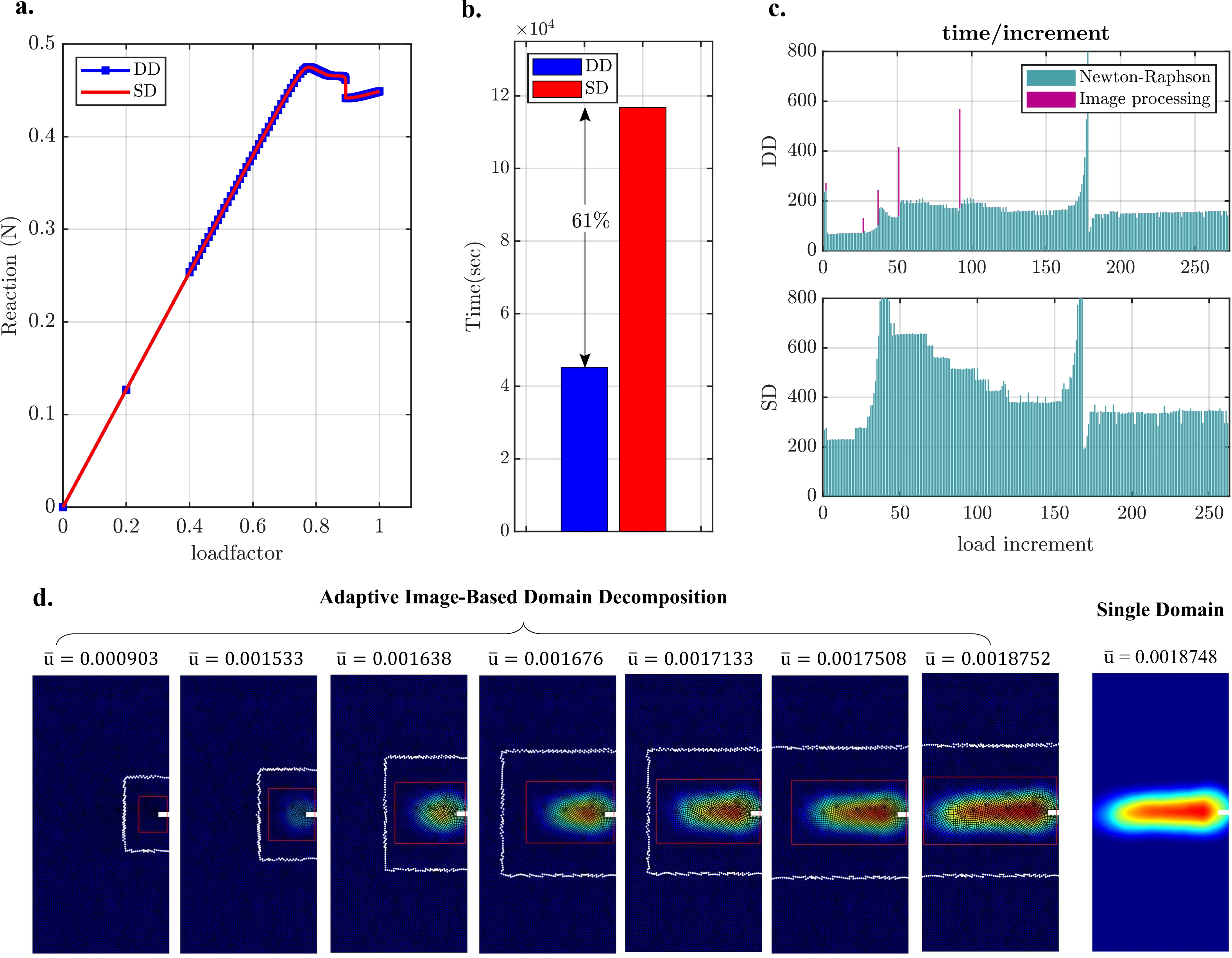}
    \caption{Numerical results for the $SNT_{unstruct}$ model. Comparison between DD and SD analysis, showing in {\bf{a.}} reaction-loadfactor response, {\bf{b.}} total simulation times and {\bf{c.}} time/increment. {\bf{d.}} Snapshots of damage propagation using our adaptive domain decomposition framework.}
    \label{Figure_SNTunstruct_results}
\end{figure} 

Fig. \ref{Figure_SNTunstruct_results}a displays the reaction curves for DD vs SD comparison, and we observe an almost exact coincidence of the equilibrium paths that are obtained with the two methods. The total simulation times are compared in Fig. \ref{Figure_SNTunstruct_results}b, and we report that DD outperforms the SD analysis by $61\%$. A time/increment comparison is shown in Fig. \ref{Figure_SNTunstruct_results}c, which sheds additional light into the performance of the two solvers during the analysis. Compared to the $SNT_{struct}$ model, the time occupied by the image-based component in this model represents a much smaller fraction of the total time; the latter quantity is governed almost entirely by the non-linear FEM operations. Evidently, this shows that as the mesh resolution increases, the overhead imposed by the image-based operations becomes negligible. Finally, we demonstrate how our framework handles the unstructured mesh by presenting in Fig. \ref{Figure_SNTunstruct_results}d a series of damage snapshots. Here we emphasize that the damage contours still maintain their rectangular shape, but the interface (white markers) comprises nodes which are arranged in an irregular fashion. This holds true regardless the level of applied load, thus preserving the ability of our framework to conduct the entire analysis. Also, as an other sanity check of the accuracy of our approach, we present in the last graph of Fig. \ref{Figure_SNTunstruct_results}d the damage contour obtained at approximately the same level of applied load as the last DD contour. We observe an almost identical match between the two images, which further verifies the quality of our framework computations.

\subsection{Symmetric Double Notch under Tension (SDNT)}
\label{Sec:SDNT}

\begin{figure}[h!]
    \centering
    \includegraphics[width=0.89\textwidth]{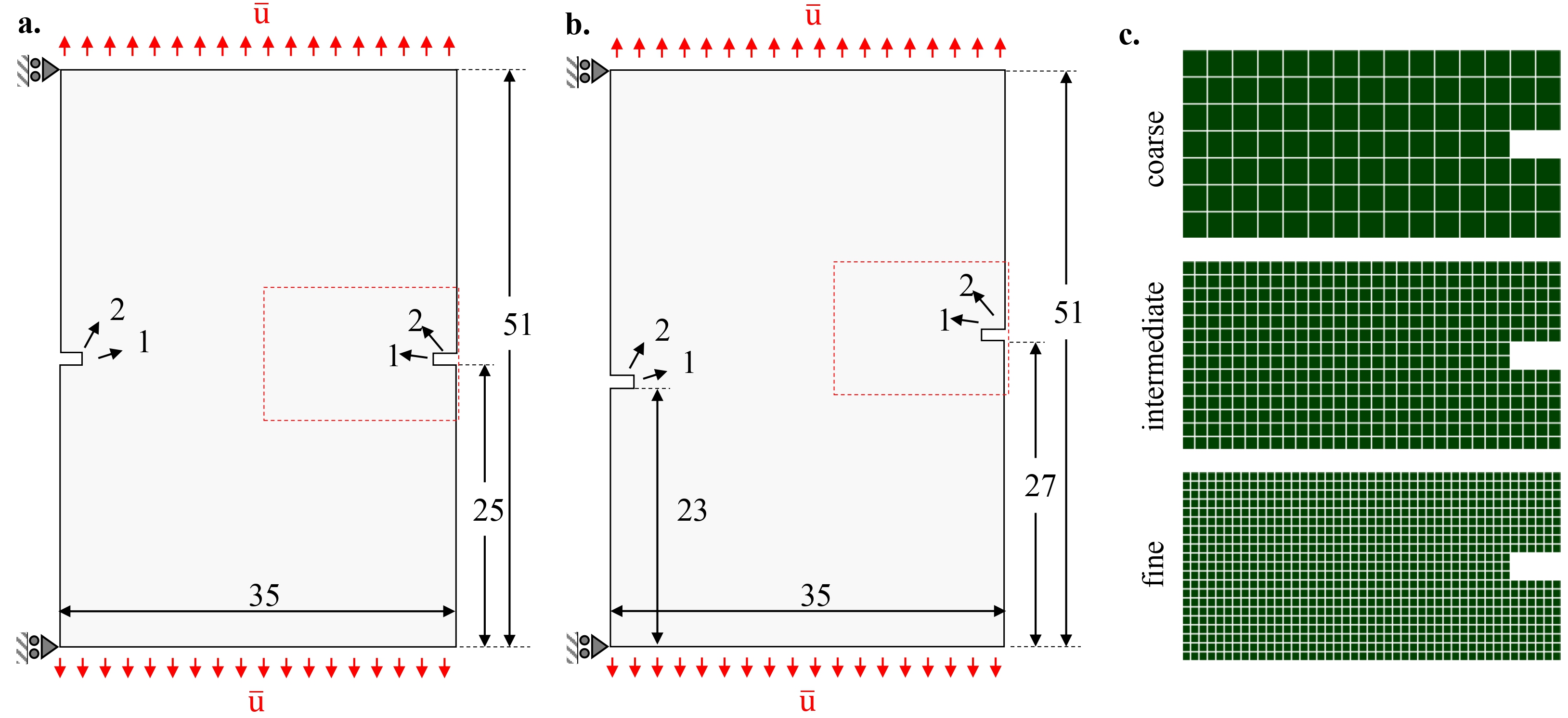}
    \caption{Geometric and loading details of the Symmetric Double Notch Tension ({\bf{a}}) and Asymmetric Double Notch Tension ({\bf{b}}) models. The same discretization is used in both cases, and the FEM mesh of the area enclosed by the red line is shown in ({\bf{c}}).}
    \label{Figure_DNT}
\end{figure}

The next numerical model is a Symmetric Double Notch under Tension (SDNT) problem. Fig. \ref{Figure_DNT}a shows a schematic view of its geometry and loading conditions. This model has a structured mesh, and we apply the non-local integral damage law and the equivalent strain definition of Eqn. \ref{Eqn_eqstrain_shear}. The goal of this variation is to showcase that our framework is not restricted to cases with just a single damage zone, but instead it can identify and capture the propagation of multiple damage paths inside the domain. Consequently, in cases where multiple damage zones get in progressively closer proximity, we show that our domain decomposition algorithm is also flexible enough to merge them into a single damage zone. For the SDNT problem we use three mesh discretizations, as shown in Fig. \ref{Figure_DNT}c. The models are termed Coarse, Intermediate and Fine, and they have 1781, 7124 and 16029 finite elements respectively. 

\begin{figure}[H]
    \centering
    \includegraphics[width=0.95\textwidth]{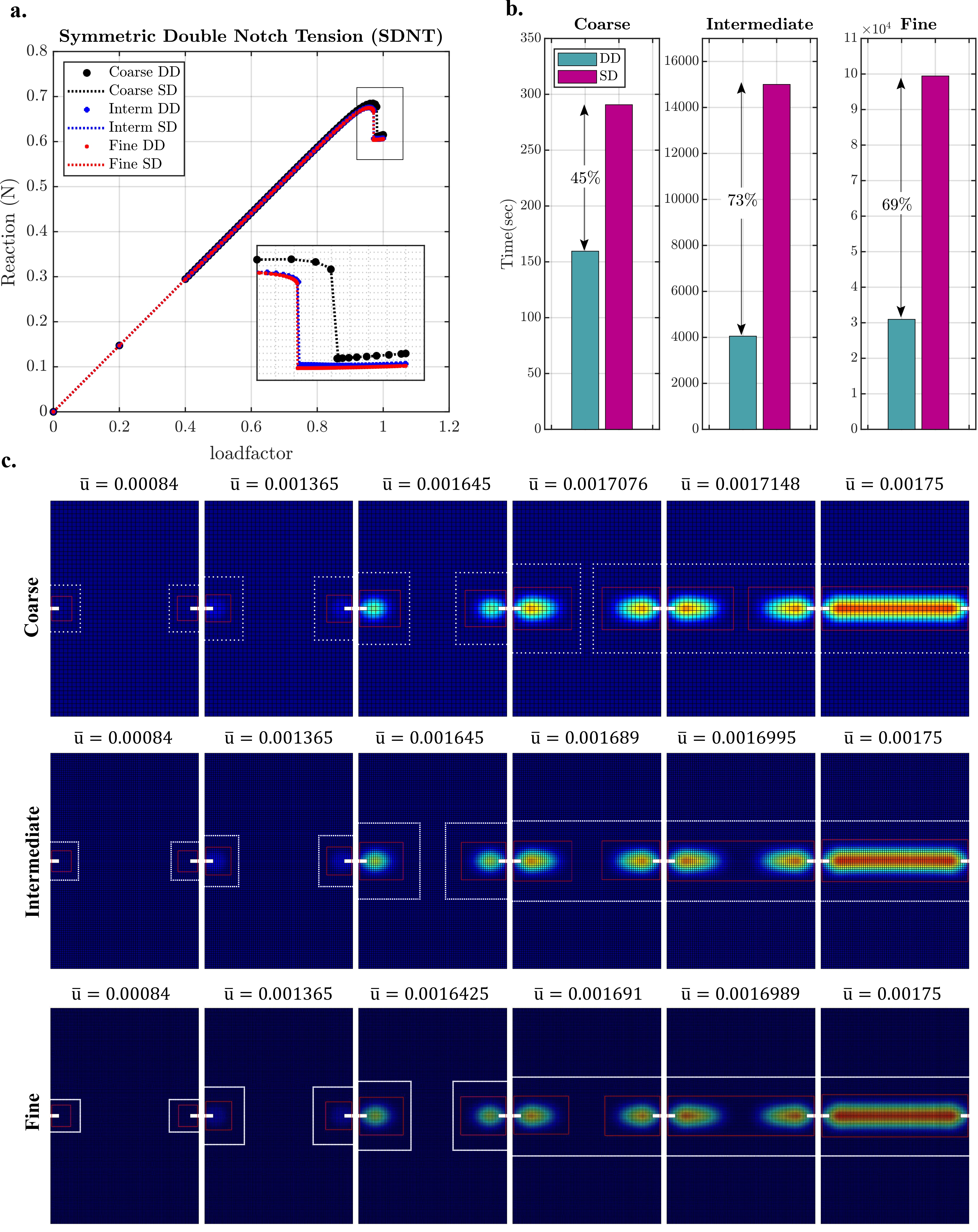}
    \caption{Numerical results of the SDNT model: {\bf{a.}} reaction-loadfactor curves, {\bf{b.}} total simulation times, {\bf{c.}} damage snapshots.}
    \label{Figure_SDNTresults}
\end{figure}

Fig. \ref{Figure_SDNTresults}a and \ref{Figure_SDNTresults}b present the reaction curves and total simulation times respectively, for both DD and SD. Similar conclusions as before are drawn, further verifying the applicability and very good performance of our framework. Also, snapshots of the damage propagation for the three meshes are shown in Fig. \ref{Figure_SDNTresults}c. This figure shows clearly: a) the ability of our image detection module to capture multiple regions of damage, and b) the ability of our domain decomposition module to perform sequential splitting and merging of the healthy and unhealthy domains. These features are present regardless of the level of mesh discretization. Overall, these functionalities enhance the versatility and robustness of our method, enabling it to model the more challenging case of damage propagating from multiple fronts.

\subsection{Asymmetric Double Notch under Tension (ADNT)}
\label{Sec:ADNT}

\begin{figure}[b!]
    \centering
    \includegraphics[width=0.95\textwidth]{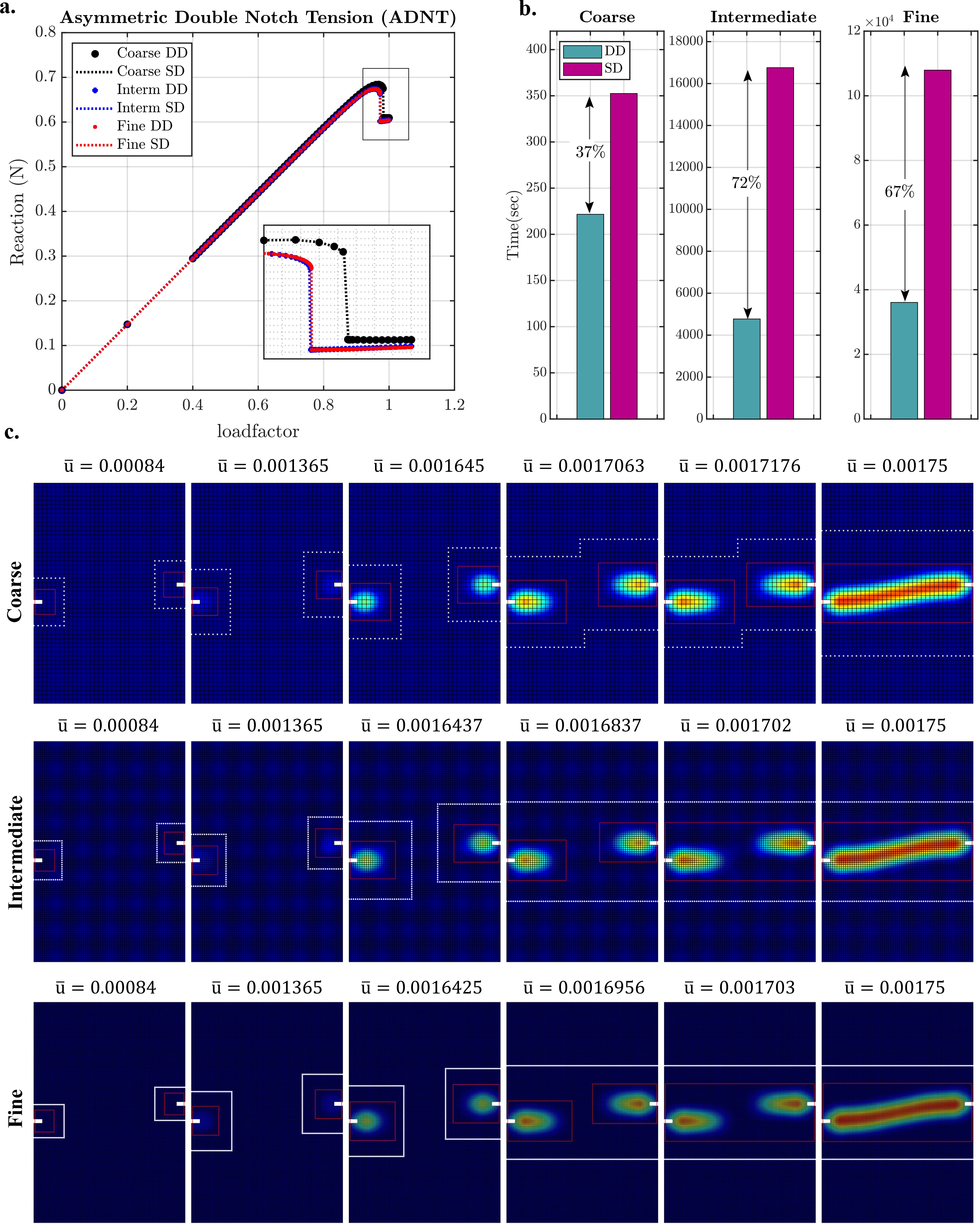}
    \caption{Numerical results of the ADNT problem: {\bf{a.}} Reaction-loadfactor curves, {\bf{b.}} total simulation times, {\bf{c.}} damage snapshots.}
    \label{Figure_ADNTresults}
\end{figure}

The last numerical model is the Asymmetric Double Notch under Tension (ADNT) problem. A schematic of this model geometry and boundary conditions is shown in Fig. \ref{Figure_DNT}b. This problem is modeled with a structured mesh, the non-local integral damage model, and the equivalent strain definition of Eqn. \ref{Eqn_eqstrain_shear}. Here we demonstrate additional capabilities of our framework, showing that our algorithm can handle asymmetries in the interfaces of the several damage zones, as well as its ability to capture the formation of diagonal damage paths. The same mesh discretizations as in SDNT are utilized, shown in Fig. \ref{Figure_DNT}c, which result in a Coarse, Intermediate and Fine mesh idealization.

The results of this investigation are plotted in Fig. \ref{Figure_ADNTresults}, which follows an identical layout as Fig. \ref{Figure_SDNTresults}. Overall, similar notes to the SDNT problem can be drawn, which further validates the accuracy of our framework. Regarding its computational performance, we report in Fig. \ref{Figure_ADNTresults}b computational savings in the order of $37\% - 72\%$ compared to SD. Finally, in Fig. 
\ref{Figure_ADNTresults}c we show snapshots of damage propagation across the three models and at various levels of external load. These graphs illustrate in action the ability of our framework to handle asymmetries in the interface locations and to trace the formation of slanted damage paths, while maintaining its ability to split or merge its subdomains.

\section{Summary and Conclusions}
\label{Sec:Summary_Conclusions}

The main novelty of this article is the development of an image-based adaptive domain decomposition framework to aid the solution of continuous damage mechanics problems within the FEM scheme. The framework consists of four modules: a) FEM numerical solver, b) contour detection algorithm, c) damage tracking sub-routine, and d) domain decomposition algorithm. A series of image-processing steps leads to the detection of damage, and then the domain is divided to a healthy subdomain (only undamaged elements) and unhealthy subdomains (damaged and undamaged elements). The robustness and versatility of our method are extensively demonstrated through several numerical examples, and the main functionalities of the framework are summarized below: 

\begin{itemize}

    \item Our approach can be used to model both local and non-local damage evolution. In this work we adopted the non-local integral model. 
    
    \item The method is insensitive to the choice of finite element mesh (size, shape and orientation), being compatible with both structured and unstructured mesh idealizations.
    
    \item For the problems considered, we report computational savings up to $73\%$ faster than a single-domain analysis, without compromising the numerical accuracy. 
   
    \item Our framework can capture the presence and monitor the concurrent propagation of multiple damage propagation paths. Also, it is flexible enough to adaptively grow or merge multiple unhealthy FEM zones, depending on the extent of damage propagation.
    
    \item The method can capture the formation of both straight or diagonal damage paths, which renders our approach suitable for a wide range of damage mechanics problems.
    
    \item Our method is insensitive to the user-defined colormap, and thus it is compatible with several software packages that may use problem-specific printing settings.
    
\end{itemize}

Finally, we note that this work can easily transition beyond the damage mechanics world and it can be directly extended to other non-linear models, such as plasticity and phase-field. Our ongoing work focuses on extending these ideas to the much more challenging case of 3D problems. In order to facilitate the extension of our framework to other models and methods, as well as to ensure reproducibility of our results, all the code and data used in this study will be made openly available upon publication of the article.

\section*{Acknowledgements}
\label{Section:Acknowledgements}

This work was partially supported by the Sand Hazards and Opportunities for Resilience, Energy, and Sustainability (SHORES) Center, funded by Tamkeen under the NYUAD Research Institute Award CG013. The authors would also like to acknowledge the support of the NYUAD Center for Research Computing for providing resources, services, and staff expertise.

\section*{Data availability}
\label{Section:Data_Availability}

All the code and data used in this work will be made publicly available upon publication of the article.

\newpage
\bibliography{bibliography}

\newpage
\appendix
\section{Constitutive Modeling}
\label{Appendix:Constitutive_Model}

In this study we employ two definitions for the equivalent strain. The first one is given as \cite{lemaitre2006engineering}:

\begin{equation}
    \varepsilon_{eq} = \sqrt{\langle\varepsilon_{I}\rangle^{2} + \langle \varepsilon_{II} \rangle^{2} + \langle \varepsilon_{III} \rangle^{2}}
\label{Eqn_eqstrain_tensile}
\end{equation}

\noindent where $\varepsilon_{I}$, $\varepsilon_{II}$, and $\varepsilon_{III}$ are the principal strains and $\langle \, \cdot \: \rangle$ are the Macaulay brackets.

The second definition of equivalent strain is given as \cite{pantidis2023integrated}: 

\begin{equation}
    \varepsilon_{eq} = \frac{k-1}{2k(1-2\nu)} + \frac{1}{2k} \sqrt{\frac{(k-1)^2}{(1-2\nu)^{2}}I_{1}^{2} + \frac{2k}{(1+\nu)^2}J_{2}}
\label{Eqn_eqstrain_shear}
\end{equation}

\noindent where $I_{1} = tr(\boldsymbol{\varepsilon})$ and $J_{2} = 3tr(\boldsymbol{\varepsilon} \cdot \boldsymbol{\varepsilon}) - tr^{2}(\boldsymbol{\varepsilon})$ are the strain invariants.

The phenomenological damage law by Mazars for brittle materials is used to model the damage variable $d$, where $d$ is a function of the equivalent strain $\varepsilon_{eq}$. The damage variable $d(\varepsilon_{eq})$ is defined as:

\begin{equation}
    d(\epsilon^*) =
    \begin{cases}
    0 & \text{if } \epsilon^* < \epsilon_d \\
    1-\left[\frac{\epsilon_{d}(1-\alpha)}{\epsilon^{*}}+\frac{\alpha}{exp(\beta(\epsilon^{*}-\epsilon_{d}))}\right] & \text{if } \epsilon^* \geq \epsilon_d   
    \end{cases}
\label{Eqn_Mazars}
\end{equation}

\noindent where $\epsilon_{d}$ is the threshold strain above which damage initiates. $\alpha$ and $\beta$ are material parameters obtained through experiments.

\end{document}